\documentclass[format=acmsmall,screen=true,review=false,natbib=true,nonacm]{acmart} 

\usepackage{physics}
\usepackage{array}
\usepackage{amsmath}
\usepackage{amsthm}
\usepackage{xcolor} 

\begin{document}

\title{Applications of Quantum Machine Learning for Quantitative Finance}

\author{Piotr Mironowicz}
\email{piotr.mironowicz@gmail.com}
\orcid{0000-0003-4122-5372}
\affiliation{%
  \institution{Department of Algorithms and Systems Modelling, Faculty of Electronics, Telecommunications and Informatics, Gdańsk University of Technology}
  \city{Narutowicza 11/12, Gdańsk, 80-233}
  \country{Poland}
}
\affiliation{%
  \institution{Department of Physics, Stockholm University}
  \city{Roslagstullsbacken 21, Stockholm, 114 21}
  \country{Sweden}
}
\affiliation{%
  \institution{International Centre for Theory of Quantum Technologies, University of Gdańsk}
  \city{Jana Bażyńskiego 1A, Gdańsk, 80-309}
  \country{Poland}
}

\author{Akshata Shenoy H.}
\affiliation{%
  \institution{International Centre for Theory of Quantum Technologies, University of Gdańsk}
  \city{Jana Bażyńskiego 1A, Gdańsk, 80-309}
  \country{Poland}
}

\author{Antonio Mandarino}
\orcid{0000-0003-3745-5204}
\email{antonio.mandarino.work@gmail.com}
\affiliation{%
  \institution{International Centre for Theory of Quantum Technologies, University of Gdańsk}
  \city{Jana Bażyńskiego 1A, Gdańsk, 80-309}
  \country{Poland}
}
\affiliation{%
  \institution{Department of Physics Aldo Pontremoli, University of Milan}
  \city{Via Celoria 16, 20133 Milan}
  \country{Italy}
}

\author{A. Ege Yilmaz}
\affiliation{%
 \institution{Hochschule Luzern, Institut für Finanzdienstleistungen Zug IFZ}
 \city{Suurstoffi 1, 6343 Rotkreuz}
 \country{Switzerland}}

\author{Thomas Ankenbrand}
\affiliation{%
 \institution{Hochschule Luzern, Institut für Finanzdienstleistungen Zug IFZ}
 \city{Suurstoffi 1, 6343 Rotkreuz}
 \country{Switzerland}}

\renewcommand{\shortauthors}{Mironowicz et al.}

\begin{abstract}
  Machine learning and quantum machine learning (QML) have gained significant importance, as they offer powerful tools for tackling complex computational problems across various domains. 
  This work gives an extensive overview of QML's uses in quantitative finance, an important discipline in the financial industry. 
  We examine the connection between quantum computing and machine learning in financial applications, spanning a range of use cases including fraud detection, 
  underwriting, Value-at-Risk, stock market prediction, portfolio optimization, and option pricing by overviewing the corpus of literature concerning various financial subdomains.
\end{abstract}




\maketitle

\section{Introduction}

In recent years, the convergence of quantum computing (QC) and machine learning (ML) has sparked significant interest across various domains, with finance being no exception. Quantum machine learning (QML) is a branch of ML that harnesses the principles of quantum mechanics to process and analyze data more efficiently. By leveraging the unique properties of quantum systems, such as superposition and entanglement, QML techniques have the potential to improve traditional methodologies in finance. These techniques promise to enhance predictive capabilities in areas such as portfolio optimization, market prediction, trading, pricing, and risk management. As quantum computers continue to evolve and become more accessible, the integration of QML into finance applications is expected.

This review paper delves into this burgeoning field of QML in finance, offering insights into its diverse technologies, applications, and potential implications. The paper is structured into several sections, each focusing on different facets of this intersection. Firstly, in sec.~\ref{sec:QuantitativeFinanceUseCases}, we explore various quantitative finance use cases where QML techniques can be applied, including portfolio optimization, market prediction and trading, pricing, and risk management. Subsequently, we delve into the underlying algorithms and ML methodologies that form the basis for these applications. 
Following this, the paper delves into the realm of QC and algorithms, providing a foundational understanding of QC principles and covering essential topics such as quantum gates, qubits, quantum circuits, and quantum algorithms.

The crucial part of the review is the comprehensive literature review given in sec.~\ref{sec:lit_rev}, which examines the existing body of research on various aspects of QML in finance. Within this part, we explore specific applications, including portfolio optimization in sec.~\ref{ssec:PortfolioOptimizationAndQuantumMachineLearning}, market prediction and trading strategies in sec.~\ref{ssec:MarketPredictionAndTradingStrategiesUsingQuantumMachineLearning}, pricing in sec.~\ref{ssec:DerivativePricingQML}, risk management in sec.~\ref{ssec:RiskAssessmentUnderwritingQML}, shedding light on the current state of the art and potential future directions. By providing a holistic overview of the field, this review aims to stimulate further research and facilitate advancements at the intersection of QC, ML, and quantitative finance.
Whereas our review concentrates on QML, there exist other reviews concerning QC in finance~\cite{bouland2020prospects,naik2023portfolio}.

\section{Quantitative Finance Use Cases}
\label{sec:QuantitativeFinanceUseCases}

This section provides an overview of some prominent applications in quantitative finance, including portfolio optimization in~\ref{sec:intro:portfolio}, market prediction and trading in~\ref{sec:intro:stock_prediction}, pricing in~\ref{sec:intro:option}, and risk management in~\ref{sec:intro:risk}. Each of these areas plays a vital role in finance, and can at best benefit from the use of quantum computing.

\subsection{Portfolio Optimization}
\label{sec:intro:portfolio}

The science of building an investment portfolio to maximize returns while lowering risk is known as portfolio optimization. To achieve the best possible balance between risk and reward, a complex process of carefully choosing a wide variety of assets, including stocks, bonds, and other financial instruments, is involved. A key tenet of this strategy is diversification, which distributes assets across a variety of (uncorrelated or low correlated) asset classes and industries, thereby lowering the portfolio's total risk. \cite{gunjan2023brief}

To assess the risk-adjusted return of an investment or portfolio, the \textit{Sharpe Ratio} is a commonly used metric. It is computed by dividing the excess return of the investment (or portfolio) relative to the risk-free rate by the standard deviation of the investment's returns. It was created by William F. Sharpe~\cite{sharpe1966mutual}. The formula for the Sharpe Ratio is as follows:
\begin{equation}
	r_\text{Sharpe} \equiv \frac{{R_p - R_f}}{{\sigma_p}},
\end{equation}
where $R_p$ is the expected portfolio return, $R_f$ is the risk-free rate, and $\sigma_p$ is the standard deviation of portfolio returns. A higher Sharpe Ratio indicates a better risk-adjusted return, as it shows how much excess return an investor receives for the extra volatility taken on compared to a risk-free asset. Thus, the Sharpe Ratio plays a crucial role in helping investors make informed decisions about their investments~\cite{sharpe1998sharpe,SharpeInvestopedia}.

An all-encompassing approach to portfolio optimization ought to take crucial elements into account such as transaction costs, liquidity, and constraints set by the investor. For example, the ease with which an asset can be purchased or sold without significantly affecting its price is referred to as \textit{liquidity}. Transaction costs comprise a range of expenses related to the purchase and sale of shares, such as taxes and brokerage fees. Furthermore, restrictions might be applied, limiting particular investment kinds or setting particular risk levels based on investor preferences. There are several approaches to achieving an optimal portfolio, each with special characteristics, including risk parity, factor-based models, and mean-variance optimization.

The so-called \textit{mean-variance optimization} was pioneered by Harry Markowitz in the 1950s~\cite{markowits1952portfolio}. This approach entails estimating the expected return and risk (volatility) of each asset based on historical data. Subsequently, the investor constructs what is termed as an \textit{efficient frontier}, encompassing all possible combinations of assets that offer the highest expected return for a given level of risk. To be more precise, this frontier comprises portfolios that meet the unique condition of having no other portfolio with a higher expected return given the same standard deviation of return. The optimal portfolio, in turn, finds its place along this efficient frontier, and its identification relies on determining the combination of assets that delivers the highest return for the investor's specific level of risk tolerance~\cite{bodnar2009econometrical}. Implementing the Markowitz mean-variance theory entails estimating the means and covariances of asset returns. Typically, these estimations involve using a sample mean vector and a sample covariance matrix~\cite{disatnik2012portfolio}.

There are two main types of portfolio optimization problems: unconstrained and constrained~\cite{behr2013portfolio}. The main difference between unconstrained and constrained portfolio optimization problems is the presence of constraints on certain \textit{weights} parameters in the latter. An unconstrained portfolio optimization problem is a mathematical problem in which the goal is to find the optimal portfolio weights that will maximize or minimize a certain objective function, such as the expected return or the variance of the portfolio, without any constraints on the weights. This means that the weights can take on any value, including negative values, usually meaning a short position on that asset, as long as they result in the optimal portfolio~\cite{karatzas1987optimal}. The constraints in constrained portfolio optimization problem can take many forms, such as limiting the maximum weight of any individual asset, requiring that the weights sum to 1 (to ensure that the portfolio is fully invested), or requiring that the portfolio meet certain regulatory requirements~\cite{cvitanic1992convex}. Constrained portfolio optimization is used more often in practice since it allows for realistic and practical investment scenarios. For example, it may not be realistic to allow for negative weights and to invest more than 100\% of the portfolio in a single asset.
An example of portfolio optimization of $n$ assets is the following problem:
\begin{align} 
	\label{eq:portfolioOptimization}
	\begin{split}
		\text{minimize } &\null q \cdot x^\top Q x - \mu^\top x \\
		\text{over } & x \in \{ 0,1 \}^n \\
		\text{subject to } &\null 1^\top x = B,
	\end{split}
\end{align}
where $q > 0$ reflects the risk appetite of the decision maker, $Q$ is an $n$ by $n$ real matrix specifying covariances between asset returns, $\mu$ is a vector with the expected return for each of the assets, and $B$ is the number of assets to be selected that can be interpreted as a budget. The pivotal operation entails computing the inverse of the covariance matrix and solving the associated quadratic optimization problem.

Factor-based models present another popular approach to portfolio construction. These models are utilized by investors to estimate the riskiness and relationships between securities in a portfolio. These models take into account various factors, including value, size, momentum, and quality. By incorporating these factors into the optimization process, investors aim to capture additional explanations of return beyond those explained by traditional asset classes.

An alternate approach to portfolio optimization is provided by the \textit{risk parity}~\cite{roncalli2013introduction}, whose main objective is to distribute risk equally among diverse asset classes.

\textit{Online portfolio optimization} is a dynamic approach to investment management that takes into account new information as it becomes available. This is in contrast to traditional portfolio optimization, which assumes that all relevant information is known at the time the portfolio is created. Online portfolio optimization algorithms can adjust the portfolio weights in real-time, allowing for more efficient and effective risk management~\cite{helmbold1998line,li2014online}.
\textit{Portfolio rebalancing} is the process of realigning the weightings of a portfolio of investments. It involves periodically buying or selling assets in the portfolio to maintain a desired asset allocation.

\subsection{Market Prediction and Trading}
\label{sec:intro:stock_prediction}

Market prediction refers to the process of forecasting future trends and movements in financial markets. It involves analyzing various factors such as historical data, economic indicators, market sentiment, and other relevant information to make predictions about the direction of prices, asset values, and overall market conditions. Market prediction include fundamental analysis, technical analysis, quantitative modeling, sentiment analysis, and microstructure analysis~\cite{elliott2013handbook,nti2020systematic}.
\textit{Fundamental analysis} involves examining the underlying factors that influence the value of an asset or market, by considering economic indicators, industry trends, company financials, and other relevant information to assess the intrinsic value of an asset~\cite{bauman1996review,wafi2015fundamental}. \textit{Technical analysis} focuses on studying historical price patterns and market trends to predict future price movements with the aid of tools such as statistical models, indicators, and charts, used to identify patterns that suggest potential selling or buying opportunities~\cite{nazario2017literature}. \textit{Quantitative modeling} involves using mathematical models and statistical techniques to analyze datasets and identify patterns or relationships between variables sometimes with support of AI methods~\cite{sharma2018quantitative}. \textit{Sentiment analysis} is a technique that aims to gauge market sentiment or investor emotions by analyzing news articles, social media posts, and other sources of information to monitor the overall sentiment, to assess whether investors are optimistic or pessimistic about the market~\cite{kearney2014textual,checkley2017hasty}.
\textit{Microstructure analysis} delves into the intricate details of how orders are executed, prices are formed, and liquidity is provided in financial markets~\cite{kissell2020algorithmic}. By studying the dynamics of order flow, market impact, and price discovery processes, quantitative analysts can gain insights into market behavior and optimize their trading strategies accordingly. Understanding market microstructure is crucial for developing effective trading algorithms that can navigate the complexities of modern financial markets~\cite{fastercapital-microstructure}.

\textit{Algorithmic trading} involves the use of computer algorithms to execute trading strategies at a speed and frequency that is often impossible for human traders~\cite{cartea2015algorithmic}. This approach relies on quantitative models to identify trading opportunities based on various factors such as price movements, volume, and market data. Algorithms can be designed to execute trades automatically based on predefined criteria, allowing for rapid decision-making and execution in the financial markets~\cite{chan2013algorithmic,kissell2020algorithmic}.
Market prediction is inherently uncertain and subject to various risks and limitations, as financial markets are influenced by a multitude of factors, including economic conditions, geopolitical events, investor behavior, and unforeseen events which can lead to unexpected market movements that may deviate from predictions. To improve the accuracy of the prediction of the stock market, neural networks (NNs) have been employed~\cite{boyacioglu2010adaptive,atsalakis2009forecasting,adhikari2014combination}.

\subsection{Pricing}
\label{sec:intro:option}

Options are financial derivative contracts that give the right, without imposing an obligation, to either purchase, known as a \textit{call option}, or sell, known as a \textit{put option}, an underlying asset at an agreed-upon price called the \textit{strike}, within a specified timeframe known as the \textit{exercise window}~\cite{hull2016options}. There are several types of options based on their exercise characteristics and settlement terms. The major types include:
\begin{enumerate}
	\item \textit{American} options, which can be exercised at any time before expiration, making them more flexible than European options;
	\item \textit{European} options, which can only be exercised at expiration, which simplifies their pricing compared to American options;
	\item \textit{Bermudan} options, which can be exercised at a set (always discretely spaced) number of times, what makes it intermediate between American and European options;
	\item \textit{Asian} options, or average value options, which have a payoff based on the average price of the underlying asset over a specific period rather than just the spot price at expiration;
	\item \textit{Barrier} options, which come into existence or cease to exist when the underlying asset’s price reaches a predetermined barrier level~\cite{Chen2022}.
\end{enumerate}
Optimal stopping theory deals with determining the optimal moment to "stop" or take an action to maximize the expected reward~\cite{wald2004sequential,shiryaev2007optimal}. For instance, American options can be seen as a super-martingale hedging problem for the seller and a stochastic optimal halting problem for the buyer~\cite{Foellmer2016}.

Because the characteristics that define an option are stochastic, it is often difficult to determine its fair value; for this reason, option pricing often requires the use of numerical methods. Because of its adaptability and efficient handling of stochastic parameters, Monte Carlo simulation is one of the most popular techniques~\cite{boyle1977options}. Monte Carlo simulation is used to simulate possible future outcomes by generating random samples and is particularly useful for complex option pricing problems with multiple sources of uncertainty~\cite{Qiuyi2009}. Nonetheless, for complex options Monte Carlo methods usually require significant computational resources to provide accurate option price estimates~\cite{stamatopoulos2020option}. Scientific research fields in option pricing concentrate on refining current models, creating new pricing techniques, investigating alternative market dynamics hypotheses, and advancing options trading risk management strategies. The goal of \textit{stochastic volatility models}~\cite{ball1994stochastic} is to increase option pricing accuracy~\cite{fastercapital-volatility} and better represent the dynamics of volatility in financial markets. \textit{Jump diffusion models} incorporate sudden jumps in asset prices into the modeling framework to account for extreme market events that impact option prices~\cite{del2015effects}.

One of the most well-known methods for option pricing is the \textit{Black-Scholes model}. By taking into account variables such as the price of the underlying asset, the strike price, the time to expiration, the risk-free rate, and volatility, it offers an estimate of the value of European-style options~\cite{black1973pricing}.
The model assumes that the stock price follows a lognormal distribution, and stock returns a normal distribution; there are no transaction costs or taxes; the risk-free interest rate is constant; and the option can only be exercised at expiration; there is no arbitrage nor dividends~\cite{corporatefinanceinstitute-BS}.
Let us denote by $C(S_t,t)$ the call option price at time $t$ and the stock price $S_t$ at $t$, $K$ is the strike price of the option, $T$ is the time of maturity of the option, $r$ is the risk-free interest rate, and $N(x)$ is the cumulative distribution function of the standard normal distribution.
The model introduces the following formulae:
\begin{subequations}
	\label{eq:BS-European}
	\begin{equation}
		\label{eq:BS-European-call}
		C(S_t,t) = S_t N(d_1) - Ke^{-r(T-t)} N(d_2),
	\end{equation}
	\begin{equation}
		\label{eq:BS-European-put}
		P(S_t,t) = Ke^{-r(T-t)} N(-d_2) - S_t N(-d_1),
	\end{equation}
\end{subequations}
where
\begin{subequations}
	\begin{equation}
		d_1 = \frac{\ln(S_t/K) + (r + \sigma^2/2)(T - t)}{\sigma \sqrt{T - t}},
	\end{equation}
	\begin{equation}
		d_2 = d_1 - \sigma \sqrt{T - t}.
	\end{equation}
\end{subequations}
The former equation~\eqref{eq:BS-European-call} is for a European call option, and the latter~\eqref{eq:BS-European-put} is for a European put option.
A more generic numerical method for the valuation of options is the binomial options pricing model. It is a lattice-based model utilizing discretized time to represent the fluctuating price of the underlying financial instrument. This approach is useful in situations when the closed-form Black-Scholes formula is not feasible. It can be applied to both American and Bermudan options since it employs the instrument's description over a period of time rather than a single point. Although it requires more computing power than the Black-Scholes model, it is often more accurate. In 1979, Rendleman and Bartter separately, and Cox, Ross, and Rubinstein (CRR) formalized it~\cite{cox1979option,Rendleman1979}.

The Heath-Jarrow-Morton (HJM) framework is a mathematical model to price interest rate derivatives~\cite{heath1992bond}. HJM and its variants are described by the following stochastic differential equation:
\begin{equation}
	df(t,T) = \alpha(t,T) \, dt + \sigma(t,T) \, dW(t)
\end{equation}
where $T$ denotes \textit{maturity}, i.e. the time at which the final payment is due on a financial instrument. $df(t,T)$ represents the instantaneous forward interest rate of a \textit{zero-coupon bond}, i.e. a bond that does not make periodic interest payments but profits to the investor from the difference between the purchase price and the face value, with maturity $T$. The parameters $\alpha(t,T)$ and $\sigma(t,T)$ are describing the \textit{drift}, i.e. systematic tendency of the stochastic process, and the \textit{diffusion} i.e. random fluctuations over time, respectively. $dW(t)$ denotes a Brownian motion under the risk-neutral assumption~\cite{HJM-Investopedia}.

\subsection{Risk Management}
\label{sec:intro:risk}


Value-at-Risk (VaR) is a widely used measure in risk management to quantify the potential loss on an asset or portfolio over a specific time horizon and with a certain confidence level. Selecting a \textit{time horizon} (such as one day or one week) and a confidence level $\alpha$ (such as $95\%$, $99\%$) are necessary for the calculation of VaR. The possible loss that could be exceeded with the selected confidence level over the given time horizon is then estimated using past data or statistical models~\cite{philippe2001value}. When the probability of suffering a loss smaller than VaR is (at least) $p$, and the likelihood of suffering a loss larger than VaR is (at most) $1-p$, then we say we have $p$ VaR. A loss that is greater than the VaR threshold is called a \textit{VaR breach}~\cite{Holton2013}.

Let us consider a portfolio with $K$ assets, where $\mathbb{L} \equiv (L_1, \cdots ,L_K)$ with $L_i \in \mathbb{R}_{+}$ denote possible losses associated with the relevant asset. The total loss is denoted as $\mathcal{L} \equiv \sum_{i \in [K]} L_i$, and its expected value is $\mathbb{E}[\mathcal{L}] = \sum_{i \in [K]} \mathbb{E}[L_i]$. Let $\alpha \in [0,1]$ be the confidence level. VaR is then defined in the following way:
\begin{equation}
	\label{eq:VaR}
	\text{VaR}_{\alpha}[\mathcal{L}] \equiv \inf_{p \in \mathbb{R}_{+}} \{ \mathbb{P}[\mathcal{L} \leq p] \geq \alpha \}.
\end{equation}
\textit{Conditional Value at Risk} (CVaR), also known as \textit{expected shortfall}, is the anticipated loss for losses exceeding VaR. CVaR is particularly responsive to extreme events occurring in the tail of the loss distribution.

The \textit{economic capital requirement} (ECR)~\cite{banks2010banking} is a critical risk metric representing the capital needed to maintain solvency at a given confidence level. It is defined as the difference between VaR and the expected loss:
\begin{equation}
	\text{ECR}_{\alpha}[\mathcal{L}] \equiv \text{VaR}_{\alpha}[\mathcal{L}] - \mathbb{E}[\mathcal{L}].
\end{equation}

A pertinent and significant issue revolves around estimating the probability of debtors reimbursing their loans, which is a crucial quantitative matter for banks. Financial institutions typically aim to gauge the creditworthiness of debtors by categorizing them into classes known as \textit{credit ratings}. These institutions have the option to develop their credit rating model or rely on credit ratings provided by major rating agencies. Borrowers are commonly divided into two primary categories based on their creditworthiness: investment-grade borrowers with low credit risk and sub-investment-grade borrowers with higher credit risk. When a borrower's rating declines from investment to sub-investment grade, they are referred to as \textit{fallen angels}~\cite{Chen2021,leclerc2023financial}.


\textit{Underwriting} is a critical process in the insurance industry where an insurer assesses the risk associated with insuring a particular individual or entity and determines the terms and conditions of the insurance policy, by analyzing various factors such as health history, lifestyle choices, occupation, and more to determine the likelihood of a claim being made~\cite{Banton2023}.

The \textit{Local Outlier Factor} (LOF) algorithm is a widely studied unsupervised anomaly detection (AD) method, frequently employed due to its effectiveness. It operates through three key steps: determining the k-distance neighborhood for each data point $x$; computing the local reachability density of $x$; and calculating the local outlier factor of x to determine its abnormality~\cite{Breunig2000,guo2023quantum}. However, it's important to note that the LOF algorithm can become computationally expensive, especially when dealing with large datasets.
\textit{Fraud Detection} (FD), a particular case of AD is a critical area of focus to safeguard financial markets, institutions, and investors from fraudulent activities~\cite{west2016intelligent,hilal2022financial}. Utilizing advanced data analytics and ML algorithms is crucial for detecting anomalies and patterns indicative of fraudulent activities in financial transactions~\cite{buanuarescu2015detecting}. Understanding the behavioral patterns of individuals or entities involved in financial transactions is essential for fraud detection~\cite{baesens2015fraud}. Network analysis involves examining the relationships and connections between different entities in the financial system to uncover potential fraud schemes~\cite{chiu2011internet}.
We refer to~\cite{app12199637} for a review of ML methods in fraud detection.

\section{Machine Learning}
\label{sec:MachineLearning}

\begin{figure}[!htbp]
        \centering
        \includegraphics[width=0.45\textwidth]{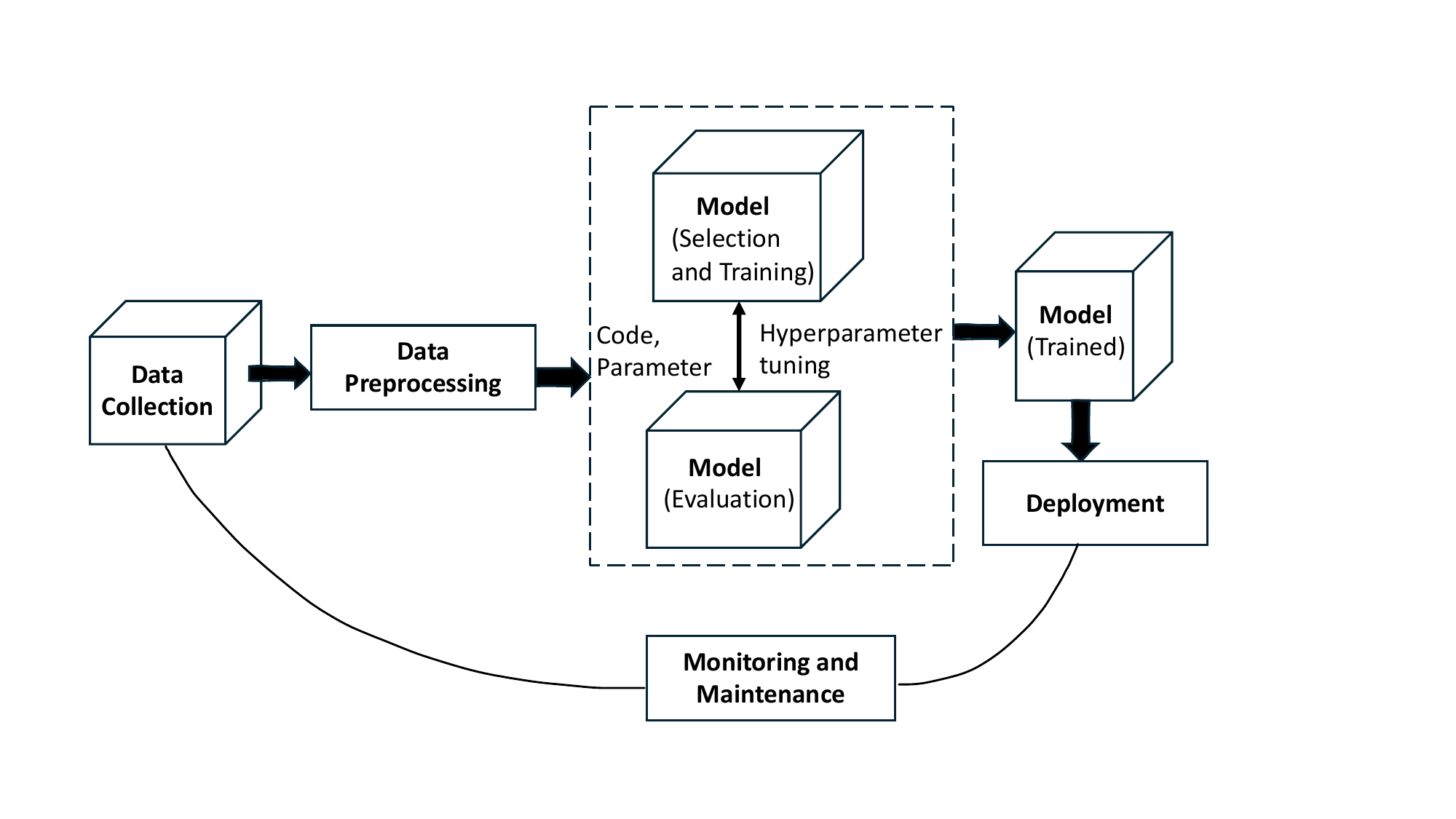}
        \caption{A general ML process is depicted here.}
        \label{fig:ML_process}
    \end{figure}

Machine Learning (ML) is an approach in the development of algorithms and statistical models enabling learning by leveraging data without explicit programming \cite{wang2009brief}. The algorithm is data-driven and hence adaptable to newer inputs improving their performance over a period. This makes them invaluable when the existence of a suitable algorithm is uncertain but abundant data is available. In the financial sector, ML has made significant impact from algorithmic trading to fraud detection, though their influence has worked both ways \cite{bundy2017preparing}. The general steps involved in training a ML algorithm (fig.(\ref{fig:ML_process})) are described as follows. 
\begin{itemize}
\item Formulation of the problem to be solved and collection of data. 
\item Preprocessing the data \cite{zelaya2019towards} for managing the missing values, normalization, dimension reduction using feature selection or feature extraction techniques. In the case of supervised learning, the processed dataset is typically split into training and testing sets.
\item Suitable model is selected and trained to predict the target outcomes while optimizing its parameters and performance. The accuracy of the model is assessed with suitable metrics. The model is evaluated on unseen data and further tuning in terms of the hyperparameters is carried out. 
\item The model is implemented in real-world scenarios with regular maintenance to ensure it is up-to-date.
\end{itemize}

Generally a dataset is comprised of multiple data points, each characterized by one or more attributes known as features \cite{van2009dimensionality}. The total number of features is the dimension of the data. Large dimensional datasets increase the computational complexity associated with training a model known as the ``Curse of Dimensionality''. This leads to the risk of overfitting and inefficiency in the training process. As a result, dimension reduction techniques namely feature selection (FS) and feature extraction (FE) emerge as key pre-processes in addressing them \cite{khalid2014survey}.

FS reduces the dimensionality of a dataset by identifying and retaining the most relevant features that contribute to the learning process. Data is not merely discarded but methodically selected to expedite training and enhance the interpretability of a model. Effective FS relies on three criteria \cite{kotsiantis2011feature}: \textit{Relevance} is the importance of a feature in making accurate predictions, \textit{Redundancy} eliminates features that do not contribute to additional information, while \textit{Diversity} ensures that the selected data includes features that provide a comprehensive representation of the unprocessed data necessary to address the task. However, if a small subset of data is selected from a very diverse collection of features, the risk of loss of information still persists. FS techniques comprise of: (i) The computationally less intense \textit{Filter methods} which rely on the statistical relation between the feature and the target outcome without any dependency on the ML algorithm. E.g. Correlation coefficient scores, Chi-Square test. (ii) The computationally expensive \textit{Wrapper methods} create a subset of features based on the ML model and achieve the best FS with respect to it. E.g. Forward selection, Backward elimination and Recursive feature selection. (iii) \textit{Embedded methods} integrate FS into the training process of a model itself and perform better than the other two methods. E.g. LASSO and RIDGE regression.

While FS chooses a subset of features from the original data, FE transforms these features into new ones while retaining all the relevant information. Often creating a subspace of lower dimension, the process removes correlations between features\cite{zebari2020comprehensive}. FE methods include: (i) \textit{Principal component analysis} where new variables known as principal components are created as a linear combination of the original feature set. Each principal component is uncorrelated with each other. E.g. Covariance matrix calculation, Eigenvalue and eigenvector computation and sorting. (ii) \textit{Linear discriminant analysis} finds a linear combination of features that maximizes the separation between different classes or groups in the dataset, useful in classification tasks. They include within-class scatter matrix, between-class scatter matrix calculation, eigenvalue and eigenvector computation to name a few. (iii) \textit{t-distributed Stochastic Neighbor Embedding} is a non-linear technique primarily used in exploratory data analysis. It maps high-dimensional datasets to a lower-dimensional space, preserving the similarity relationships between data points. E.g. Probability distribution calculation, similarity calculation and optimization gradient descent.

Finally, generalization of a model, to better capture its predictive powers on unseen data that was not used during the training process but originates from the same distribution \cite{doshi2018considerations} is discussed. A generalized model should ideally recognize irregularities and complexities present in this data without overfitting. Patterns and structures identified in the training dataset must be applied accurately on this unseen data. A simpler model is preferred over a complex one when both work equally well with the same predictiveness. Accuracy, precision, recall are certain metrics that analyze the performance of a generalized model. In general, it is known that complexity of the model, quality of the dataset employed in training and regularization techniques affect the generalization process.

ML can be broadly classified into, Supervised learning, Unsupervised learning and Reinforcement learning, based on the algorithms employed for learning. The algorithms are further characterized in terms of their interpretability and explainability to understand why some models outperform others \cite{yoo2005machine}. Interpretability refers to the effect and the cause which gave rise to a particular model and its performance. It determines a model's predictive capability, in the event a change in its input parameters occur, owing to its implementation in realistic scenarios. Higher the interpretability better is the understanding of how a model arrives at its predictions. This is especially critical for financial transactions where the stakes associated with decisions are greater. For instance, a model's incorrect decision to block a legitimate credit card transaction could have serious consequences in case of an emergency. When a model lacks interpretability, it becomes necessary to employ additional methods to make its decisions comprehendible, leading us to the concept of model explainability. Here, the focus is on elucidating the model's behavior, even if the inner workings driving its predictive powers are not entirely clear. As ML continues to evolve, balancing the performance of models to have a clear understanding of their decision capability is necessary. 

\subsection{Supervised Learning}
Supervised learning (SL) algorithms generate a function that map inputs to desired outputs using labelled datasets. They are divided into different classes of finite order, forming the basis for model training. SL model trains to employ the labelled datasets and associate the correct output to the corresponding input via mathematical analysis \cite{nasteski2017overview}. SL can be divided into classification and regression algorithms. A classification model maps its inputs to predetermined discrete outputs. Most common example is classifying emails into `spam' or `not spam' classes. A regression model maps an input to a continuous or a real-valued output. For instance, the value of a house may vary depending on its construction year, area and location. We now briefly discuss some of the SL algorithms.

\textit{Linear Regression} (LR) predicts outcomes accurately using a set of optimal coefficients. The key assumptions here are that the relationship between the input and output variables is linear (\textit{Linearity}), both the variables are error-free, input variables are independent of each other (\textit{Absence of multi-collinearity}) and normalized (\textit{Normalization}). For correlated inputs, LR tends to overfit while their normalization ensures better predictions. A general LR model can be expressed as, $ \mathcal{M}_{\beta_{0}\beta_{1}}(X) = \beta_{0}+\beta_{1} X$, where $X$ is the input variable, $\beta_{1}$ is a d-dimensional vector of coefficients and $\beta_{0}$ is a real number known as the bias coefficient or intercept. This model $\mathcal{M}_{\beta_{0}\beta_{1}}(X)$ is utilized to predict the output variables $Y$ ($Y\longleftarrow \mathcal{M}_{\beta_{0}\beta_{1}}(X)$). Different coefficients $(\beta_{0},\beta_{1})$ and $(\beta'_{0},\beta'_{1})$ lead to two different models $\mathcal{M}_{\beta_{0}\beta_{1}}(X)$ and $\mathcal{M}_{\beta'_{0}\beta'_{1}}(X)$ predicting $Y$ and $Y'$ respectively. Such models are very useful in predicting the financial and economic performances of companies, banks\cite{bakar2009applying}, stock market and trading\cite{altay2005stock}. Though a LR model is simple and easy to interpret, it is not always suitable for classification problems since it predicts continuous-valued outcomes. A more accurate approach would be to employ logistic regression where the output is binary. 

\textit{Logistic Regression} is useful for categorical output variable. An input $x$ is said to belong to a particular class $y$ or not, corresponding to $P(y=1|x)$ (positive decision) or $P(y=0|x)$ (negative decision), based on $\sigma(x)= \frac{1}{1+e^{-x}}$ known as the sigmoid function. Here, the relationship between the input and the output is not linear. The choice between linear and logistic regression depends on the nature of the target outcome. Logistic regression is very aptly suited for binary classification in high risk scenario\cite{bracke2019machine} where precise financial decisions are to be made.

\textit{Decision Trees} are models with tree-like structure for performing decision tasks\cite{navada2011overview}. They consists of nodes, branches and leaves. The basic node is known as the root and has no incoming edges. Internal nodes have outgoing edges and represent a test on the features. Each branch indicates the outcomes of this test while the leaves are a class or a continuous value. Starting from the root of the tree, data is split into different nodes based on certain criteria using the branches until the final outcome is reached. Decision trees can be categorical or regression type. They are easy to understand and do not require data normalization. Such models very useful in business decision-making tasks\cite{gepp2010business}. A major disadvantage of such algorithms are risk of overfitting and model instability for small variations in input data. 

\textit{Support Vector Machines (SVMs)} are helpful in linear and nonlinear classification problems\cite{pisner2020support}. In this formalism, an optimal hyperplane is constructed using support vectors, that clearly separates two classes of data. Support vectors are the closest data points to the hyperplane. A hyperplane is mathematically represented as  $ \beta_{0}+\beta_{1}. X=0$ where $\beta_{0}$ is the bias and $\beta_{1}$ is the $p-$dimensional vector representing a set of data points $\{(x_i,y_i)\}^{n}_{i=1}$, $n$ is the length of the sample and $y_n=\pm 1$ depending on the class to which the data point belongs to. For instance, consider the binary classification problem\cite{gaspar2012parameter}, where the aim is to construct a hyperplane that divides the datasets into two classes. All the points corresponding to $y_n=1 (-1)$ belong to the first (second) class and lie on one (other) side of the hyperplane.  Although (SVMs) are meant for general linear classification, with the help of Kernel functions, they can be used for non-linear cases as well. Kernel functions map non-linear datasets into a larger dimensional linear space which can then be separated as different classes using a hyperplane. Commonly used kernel functions are polynomial functions, radial basis functions and sigmoid functions. Support vector machines are suitable for small or medium-sized datasets of data mining and have been extensively applied for financial forecast.

\textit{Support Vector Regression (SVR)} is an extension of SVMs though they are useful for predicting continuous values as opposed to classification tasks.
Suppose $\{(x_i,y_i)\}^{n}_{i=1}$ is the available training data set\cite{awad2015support}, then the basic idea of the SVR is to find a function $f(x)$ that predicts the target values with no more than $\epsilon$ deviation from the actual values. A best fit with $\epsilon$-threshold in contrast to a best hyperplane is looked for. Represented as a tubular region around the regression line, the $\epsilon$-region encompasses the errors points which are not penalized. Any other point lying outside this region is penalized. In addition to the $\epsilon$-threshold, the function $f(x)$ is expected to be as flat as possible. The trade-off between the $\epsilon-$deviation and the flatness can be captured by the C parameter. Similar to SVMs, the input dataset can be transformed into a higher dimensional space using common Kernel functions. The most common use cases of this formalism w.r.t. finance includes stock market exchange and currency conversion\cite{henrique2018stock}.

\textit{Neural Networks (NNs)} perform computations using artificial neurons which are several interconnected nodes that mimic a human brain and its functioning. A NN comprises of an input layer for receiving the input data, an output layer for accessing the targeted output and several in between hidden layers for computation and FE depending on the complexity of the problem\cite{choi2020introduction}. Neurons in each layer process inputs received by them and pass their outputs to the next layer for further processing. Associated with each neuron is an activation function (also known as the transfer function) which defines the learning ability of a neuron. This function indicates whether a neuron has to be activated or not for predicting the targeted outcomes. Typical examples of activation functions are the sigmoid and rectified linear unit. The importance of each neuron can be determined by its weight, bias and the activation function. Upon obtaining the final output from the network in a single run, the network re-calculates these parameters for each neuron via a feedback process, to minimize error propagation and predict accurate results. A stable NN is obtained after iterating over several forward and backward propagations. NNs are broadly classified as convolutional NNs, recurrent NNs and deep learning NNs. Though primarily employed for pattern recoginition, speech recognition and natural language processing, these networks can also be used for classification problems similar to logistic regression. For instance, a sigmoid function can be employed to find a separation between classes via hyperplanes (known as the perceptron algorithm). They find application in option trading and financial forecasting.

Finally, w.r.t. classification tasks, we would like to point out some metrics useful in visualizing the performance of any model \cite{tut}. The receiver operating characteristic (ROC) curve measures the performance of a classification model for various threshold settings. The curve shows the variation of the true positive rate against the false positive rates at each threshold point. The true positive rate (false positive rate) is defined as the ratio of the true positive (false positive) to the sum of true and false positives. Also known as the sensitivity or Recall, this metric gives the model's ability to infer the proportion of true positives and false positives correctly. An efficient algorithm that computes the points under the ROC curve is known as the Area under the curve (AUC). An average of the performance across all threshold settings can be obtained here. AUC varies from 0 to 1 indicating that the model's predictions are completely incorrect or correct respectively. It is interesting to note that AUC infers the predictive capabilities of a model irrespective of threshold classification. Furthermore, it is noteworthy to mention radial basis functions for SVMs \cite{ding2021random} and NNs \cite{montazer2018radial}. These are radially symmetric functions which give the distance of an input from a chosen fixed point. This real-valued function is especially useful for introducing non-linearity in regression and classification models.

\subsection{Unsupervised Learning}
Unsupervised learning (USL) is applicable in ML for uncategorized datasets where the data inputs do not have  corresponding labelled outputs. USL algorithms discover the inherent structure in the dataset by identifying latent variables and patterns without explicit feedback or supervision, in contrast to SL. USL algorithms can be classified into clustering and Association algorithms. 

\textit{Clustering algorithm} is a fundamental USL technique where data points are organized into distinct groups (clusters) based on their similarities and differences \cite{naeem2023unsupervised}. Clustering algorithms are versatile and can be categorized as follows: \textit{Exclusive Clustering} where each data point belongs to exactly one cluster. E.g. K-Means clustering assigns data points to the nearest cluster center. In \textit{Overlapping Clustering} a data point can belong to one or multiple clusters. E.g. Fuzzy C-Means where data points have degrees of belonging to various clusters. \textit{Hierarchical Clustering} builds a hierarchy of clusters, either agglomeratively (bottom-up) or divisively (top-down) continuously merging and reorganizing them during the training process. E.g. Nested structures in datasets. \textit{Probabilistic Clustering} assigns data points to clusters based on their probability of belonging assuming different probability distributions. E.g. Gaussian mixture models.

\textit{Association Algorithms} find correlations, patterns or relationships between data points within large datasets. A combination of one or more data points that appear in the dataset known as Itemsets, meeting a minimum support threshold, are identified. Support refers to how often a data point occurs in the dataset. Association or relationship rules  (If-Then statements) are then built based on how often these itemsets occur. For instance, Market basket analysis \cite{hruschka2021comparing}, where patterns in customer purchases are identified and analyzed is an apt example. The commonly known association algorithms are the A-Priori algorithm employing bottoms-up approach in mining frequent itemsets and the FP-Growth which uses a tree-like structure. The algorithms are important for credit risk assessment where correlations between various financial behaviors and credit risk are identified.

A significant aspect of USL is generative modeling, which focuses on understanding and capturing the internal structure of data. Generative models learn the joint probability distribution of input data, as opposed to discriminative models that focus only on the conditional probability distribution. In the context of finance \cite{mohanty2021financial}, models like Generative Adversarial Networks (GANs), consisting of a generator and discriminator, generate new and artificial data samples that mimic the real data. This is invaluable in realistic financial scenarios for augmenting datasets where real data is scarce or sensitive. Similarly, Variational Autoencoders are powerful in feature learning and dimensionality reduction. They can reconstruct input data after compressing it into a lower-dimensional subspace, useful for identifying key features in complex financial datasets \cite{xie2021unsupervised}. Other considerations in USL are anamoly detection, dimensionality reduction and data processing as mentioned before.

\subsection{Reinforcement Learning}

\begin{figure}[!htbp]
        \centering
        \includegraphics[width=0.3\textwidth]{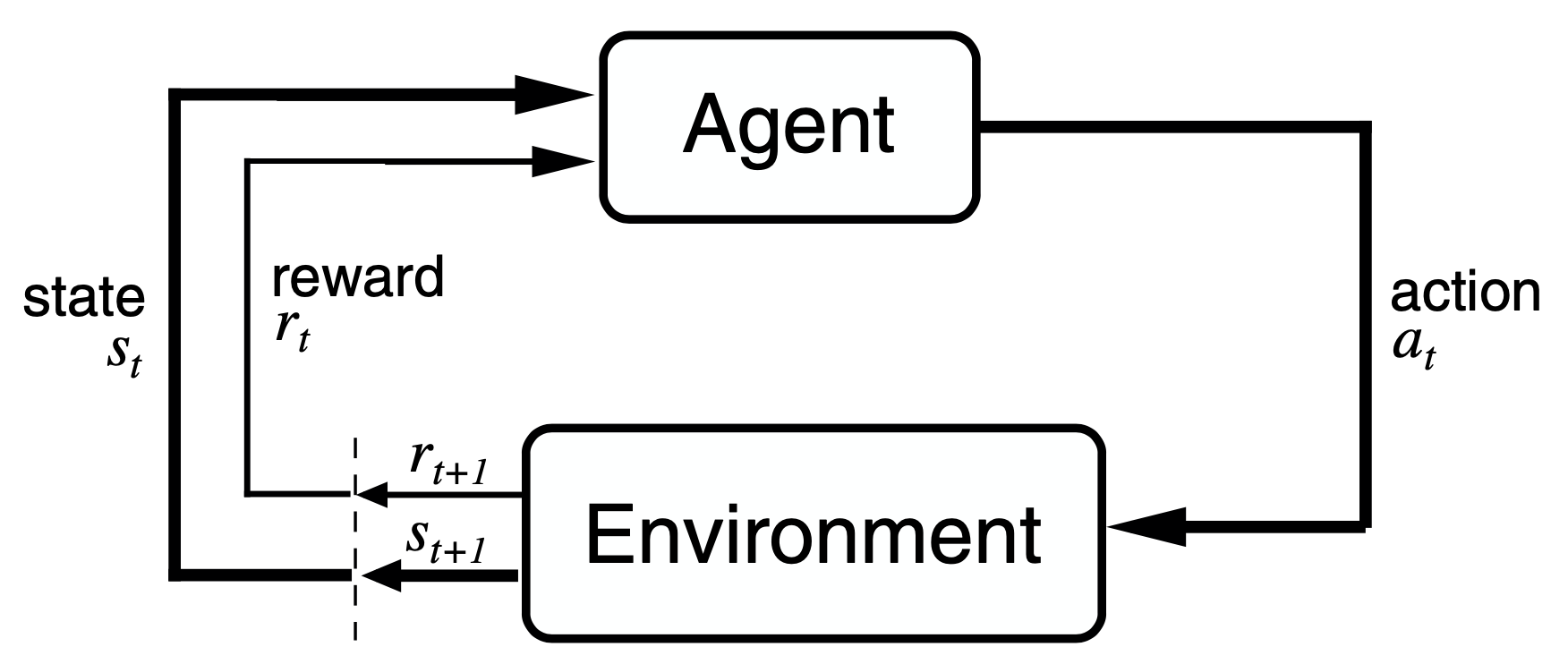}
        \caption{The transition scheme in RL. An agent acts upon receiving a response from the environment. The response contains a state $s_t$ and a reward $r_t$ at time $t$. The agent receives a new response as a result of a new action $a_t$, which consists of a state $s_{t+1}$ and a reward $r_{t+1}$. This is called a transition. Time steps could refer to arbitrary decision-making iterations, which do not necessarily have fixed intervals of real-time. Picture taken from \cite{sutton_barto}.}
        \label{fig:RL_env}
    \end{figure}

Consider the context, where a learner is expected to make decisions based on observations to solve a specific problem. Such learners in RL are called \textit{agents}. As the decisions of the agent depend on observations or \textit{states} coming from an \textit{environment}, the observations depend on the decisions or \textit{actions} of the agent. For the vast majority of the problems of practical interest, a priori knowledge of the complete specification of the environment is not available \cite{sutton_barto}, which makes classical optimization methods and supervised learning techniques inapplicable. Instead, this type of information can be revealed by interacting with the environment and developing strategies or \textit{policies} based on the responses from the environment. These responses include the state of the environment and a reward signal based on the actions taken by the agent. The process of taking an action based on an observation and receiving a response as a result is called a \textit{transition} (see Figure \ref{fig:RL_env}). RL tasks can be episodic, where transitions lead to a terminal state at a certain point and the task is restarted. They can also be continuous, where this is not the case.
    
  Some reinforcement learning algorithms are mentioned in brief here. \textit{Q-Learning} is an approach aimed to develop an optimal action policy for any Markov decision process. It is an off-policy due to random actions taken by the algorithm to gain rewards in a greedy manner. The algorithm uses a Q-function $Q(s,a)$ to estimate the value of an action $a$ taken in a particular state $s$. This helps the agent in taking further actions to maximize the cumulative reward, sometimes at the cost of sacrificing immediate gains. The Q(s,a) is updated at each step based on the actions taken, rewards assigned and the learning rate\cite{gosavi2009reinforcement} known as value iteration. The agent develops an optimal policy by making random actions known as exploitation and observing these actions to increase rewards known as exploration. Q-Learning is simple to implement and works in a model-free environment, though it cannot handle large, complex and continuous states of the environment. In such cases Deep-Q Learning, where Q-Learning is combined with deep learning is preferred\cite{arulkumaran2017deep}. The state-action-reward-state-action algorithm is another important technique which is an on-policy and hence does not make random actions similar to Q-Learning. In contrast to Q-Learning, a policy gradient algorithm learns the policy that maps each state directly to a particular action known as policy iteration. Some noteworthy hybrid algorithms that use both value based and policy based iteration approaches are Soft-Actor-Critic, Twin Delayed Deep Deterministic Policy Gradients and deep deterministic policy gradients\cite{oh2020discovering}.

\section{Quantum Computing and Algorithms}
\label{sec:QuantumComputingAndAlgorithms}

We will now cover basics of QC and QML. Interested readers seeking a comprehensive introduction to QC can refer to seminal work~\cite{divincenzo1995quantum}, and recent one~\cite{lau2022nisq}, and for QML~\cite{jhanwar2021machine}.

\subsection{Basics of Quantum Computing Methods}

Classical computers employ logic gates to execute classical algorithms, analogously a circuit-based quantum computer utilizes quantum gates to run quantum algorithms on quantum bits, known as qubits.
The mathematical description of a qubit is a vector in a two-dimensional Hilbert space $\mathcal{H}) \simeq \mathbb{C}^2.$ The Dirac notation is the standard one to denote a qubit, namely a general state can be expressed as: 
\begin{equation}
\label{eq:1qubitstate}
    \ket{\psi_1} = \alpha \ket{0} + \beta \ket{1}, 
\end{equation}
in which we have introduced the complex amplitudes $\alpha$ and $\beta$, that fulfill the normalization condition  $|\alpha|^2 + |\beta|^2 = 1$. 
The two orthonormal states $\ket{0}$ and $\ket{1}$ form the computational basis for one qubit and conventionally they are the eigenstate of the 
third Pauli matrix, such that $\sigma_3 \ket{k} = (-1)^k \ket{k}.$ 

In particular, the identity and the set of the tree Pauli matrices 
\begin{align}
\sigma_0 = \mathbb{I} =
    \begin{pmatrix}
      1&0\\
      0&1
    \end{pmatrix}, \,
\sigma_1 = \sigma_\mathrm{x} =
    \begin{pmatrix}
      0&1\\
      1&0
    \end{pmatrix}, \,
  \sigma_2 = \sigma_\mathrm{y} =
    \begin{pmatrix}
      0& -i \\
      i&0
    \end{pmatrix}, \,
  \sigma_3 = \sigma_\mathrm{z} =
    \begin{pmatrix}
      1&0\\
      0&-1
    \end{pmatrix},
    \label{eq:pauli}
\end{align}
constitute a basis in which any observable acting on a qubit can be expanded. 
The linear combination of states in Eq.\eqref{eq:1qubitstate} is known as a superposition of the two basis states, 
and the squares of the amplitudes account for the probability of detecting the corresponding state.
Namely, we have a probability $|\alpha|^2$ $(|\beta|^2)$ that the state $\ket{\psi_1}$ collapse in $\ket{0}$ $(\ket{1})$ after a projective measurement. 

Similarly, we introduce a register of $N$ qubits 
\begin{equation}
\label{eq:Nqubitstate}
    \ket{\psi_N} = \sum_{i_1, ..., i_N = 0, 1} \alpha_{i_1, ..., i_N} \ket{i_1, ..., i_N}, 
\end{equation}
for which the normalization condition is $\sum_{i_1, ..., i_N = 0, 1} |\alpha_{i_1, ..., i_N}|^2=1.$
Any logic operation on such register is performed via a quantum gate, that is formally expressed as a unitary matrix $U$ such that $U^\dag U = U U^\dag = \mathbb{I},$ 
therefore it belongs to the $\mathbf{SU}(2^N)$ Lie group. 
Through the set of operators defined in \eqref{eq:pauli}, one can define the most relevant class of single and two-qubit parametric operations, respectively: 
\begin{equation}
    \label{eq:rotations}
    R_\alpha(\theta) = e^{- i \frac{\theta}{2} \sigma_\alpha} \, \, \text{and} \,\, C_{\alpha \alpha}(\theta) = e^{- i \frac{\theta}{2} \sigma_\alpha \otimes \sigma_\alpha},
\end{equation}
where $\otimes$ denotes the tensor product \cite{mandarino2018bipartite}. 
They are generally known as rotation and controlled gates and are used to build variational, or parametrized, quantum circuits (PQCs), thanks to the possibility of adjusting the angles $\theta$ defining each gate.

Despite significant advancements in designing quantum gates to operate on qubits, a considerable gap still persists between the theoretical results achieved in quantum algorithms, and their practical implementation on quantum processing units.  In fact, at the present stage computing systems and simulators remain distant from showcasing any form of quantum advantage in addressing problems emerging outside the purely academic interests and that could impact our daily existence. This gap can be attributed in part to the current era of noisy intermediate-scale quantum (NISQ) technologies \cite{Preskill_quantum}. Quantum processing units are constrained by a limited number of qubits and significant levels of noise, including decoherence processes that spoil all the quantum resources requested for the desired speed-up \cite{Tindall_PRXQuantum, Schuld_PRXQuantum}.  
Techniques to mitigate errors and eventually correct them, such as surface codes and the engineering of logical qubits,  have been proposed \cite{RMP_QEM, Gottesman}. Here the main approach is to map a set of physical qubits into a logical qubit, often utilizing specialized network-like circuits for quantum processors composed of logical qubits. However, given the non-unique mapping questions related to the optimal embedding arise and about the scalability of such procedure on different computing architectures. 

\subsection{Basic building blocks of hybrid algorithms}

A \textit{hybrid quantum-classical algorithm} combines elements of both quantum and classical computation to solve computational problems more efficiently than classical methods alone. These algorithms leverage the strengths of quantum computing, such as superposition and entanglement, alongside classical techniques to achieve better performance. A \textit{Quantum Processing Unit} (QPU) is a hardware device designed to perform quantum computations. It consists of qubits, the fundamental units of quantum information, and is used to execute quantum algorithms. The \textit{Quantum Approximate Optimization Algorithm} (QAOA)~\cite{farhi2014quantum} is a quantum algorithm designed to solve combinatorial optimization problems by preparing a quantum state that encodes the solution to the problem and then measuring it to obtain an approximate solution. The \textit{Variational Quantum Eigensolver} (VQE)~\cite{tilly2022variational} is a quantum algorithm used to find the ground state energy of a given Hamiltonian by optimizing the parameters of a parameterized quantum circuit. A \textit{Parameterized Quantum Circuit} (PQC) is a quantum circuit with adjustable parameters that can be optimized to solve specific computational tasks. Finally, a \textit{Quantum Support Vector Machine} (QSVM)~\cite{rebentrost2014quantum} is a QML algorithm used for classification tasks, leveraging QC principles to perform efficient classification of data points into different classes.

\textit{Quantum annealing}~\cite{johnson2011quantum} (QA) is a computational technique that leverages quantum mechanical principles to solve optimization problems. In QA, a quantum system is initialized in a simple, known state and gradually evolved towards a low-energy state that represents the solution to the optimization problem. \textit{Forward annealing} refers to the process of gradually decreasing the system's temperature or energy levels from an initial high value to a low value, allowing it to explore different configurations and settle into the ground state. \textit{Reverse annealing}~\cite{ohkuwa2018reverse}, on the other hand, involves starting from the ground state and gradually increasing the temperature or energy levels to explore different configurations before reaching the desired state. \textit{Simulated quantum annealing} (SQA)~\cite{apolloni1989quantum}, also known as quantum-inspired annealing, mimics the behavior of QA using classical computing resources, often by simulating the behavior of a quantum system undergoing annealing. While not as powerful as true QA, the simulated one can still be useful for solving optimization problems more efficiently than classical methods in certain cases.

The \textit{quantum phase estimation algorithm} is a technique for estimating the phase that corresponds to an eigenvalue of a given unitary operator~\cite{kitaev1995quantum}.
\textit{Grover} provided an algorithm for unstructured search in~\cite{grover1996fast}.
A quantum algorithm for \textit{minimal search} is shown that locates the index $y$ in a table $T$ of size $N$ such that $T[y]$ is minimum with a probability of at least $1-1/2^c$, with a time complexity of $O(c \sqrt{N})$ was given in~\cite{durr1996quantum}.

The Grover searching technique was expanded upon by \textit{Quantum Amplitude Amplification and Estimation} (QAE), which was first presented in~\cite{brassard2002quantum}. This approach considers a Boolean function $\chi:X \rightarrow \{0,1\}$, where $x$ is refered to as "good" if $\chi(x)=1$ and "bad" otherwise. Consider a quantum algorithm $\mathcal{A}$, $\mathcal{A} \ket{0} = \sum_{x\in X} \alpha_x \ket{x}$, and let $a$ represent the probability that a good element is obtained if $\mathcal{A} \ket{0}$ is measured. Next, assuming that algorithm $\mathcal{A}$ makes no measurements, amplitude amplification is a procedure that enables a good $x$ to be located. Note that $\mathcal{A}$ in Grover's searching algorithm is limited to generating an equal superposition of all members of $X$, and requires that there is a known unique $x$ such that $\chi(x)=1$. Whether the value of $a$ is known in advance or not, QAE still functions. After several applications of $\mathcal{A}$ and its inverse, which is proportional to $1/\sqrt{a}$ even in the worst scenario, a "good" $x$ can be identified if the value of $a$ is known. For a wide range of search problems for which there are effective classical heuristics, the quadratic speedup can also be achieved. It is possible to estimate the value of $a$ by combining concepts from Grover's and Shor's quantum algorithms in the \textit{amplitude estimation} process. Applying QAE to the problem of \textit{approximate counting} allows one to estimate the number of $x\in X$ such that $\chi(x)=1$.

Several ideas have been put forth on how to construct quantum modular exponentiation, multipliers, and adders utilizing a set of basic quantum gates. Reversible versions of well-known classical implementations were the first circuits that were proposed~\cite{vedral1996quantum,beckman1996efficient}. Further significant developments include~\cite{draper2000addition,cuccaro2004new,van2005fast,Draper2006,Takahashi2010,Markov2012,Thapliyal2013,zhou2017quantum}, see~\cite{ruiz2017quantum,orts2020review} for an overview.

A quantum algorithm, called Harrow–Hassidim–Lloyd (HHL), for solving systems of linear equations efficiently was demonstrated in~\cite{harrow2009quantum}. For a given linear equation $A x = b$ with the unknown $x$, and $A$ with condition number $\kappa$, and given matrix $M$, the HHL computes the expectation value of $x^\dagger M x$. For small $\kappa$ it was shown that any classical algorithm requires exponentially more time than HHL. Later, the HHL method was used in the work~\cite{wiebe2012quantum}, which efficiently determined the quality of a least-squares fit over an exponentially large data set. In many instances, the algorithm could also efficiently identify a concise function that approximated the data to be fitted and bounded the approximation error. In cases where the input data consisted of pure quantum states, the algorithm was employed to provide an efficient parametric estimation of the quantum state, and therefore, it could be utilized as an alternative to full quantum-state tomography, particularly when a fault-tolerant QPU was available.

In~\cite{lloyd2014quantum} it was demonstrated that multiple copies of a quantum system, characterized by a density matrix $\rho$, could be employed to construct the unitary transformation $\exp(-i \rho t)$. Consequently, this enabled the performance of quantum principal component analysis on an unknown low-rank density matrix. This approach allowed for the determination in a quantum form of eigenvectors associated with the largest eigenvalues, accomplishing this task in exponentially less time compared to any preexisting algorithm.

A quantum algorithm is presented by the authors of~\cite{berry2017quantum} for systems of linear ordinary differential equations with constant coefficients, including possibly inhomogeneous instances. This technique delivers an exponential improvement over previous quantum methods by producing a quantum state proportional to the answer at a predetermined final time and attaining polynomial complexity in the logarithm of the inverse error. They are simulating the evolution according to the propagator using a Taylor series and encoding the simulation into a sparse, well-conditioned linear system by leveraging the HHL approach. Their method provides improved numerical stability without requiring extra hypotheses by avoiding the drawbacks of finite difference techniques. Consequently, they present a quantum algorithm for linear differential equations with a complexity of $poly(\log(1/\epsilon))$, marking a substantial exponential improvement over existing methods~\cite{berry2014high} where the overall complexity remains $poly(1/\epsilon)$ due to the inherent error introduced by the multistep method.

The Feynman-Kac formula establishes a connection between solutions of certain partial differential equations (PDEs) and Markov processes and provides a way to solve certain types of the former by relating them to the behavior of the latter. An algorithm based on variational quantum imaginary time evolution for solving the Feynman-Kac partial differential equation resulting from a multidimensional system of stochastic differential equations was proposed in~\cite{alghassi2022variational}. The correspondence between the Feynman-Kac PDEs and the Wick-rotated Schr\"{o}dinger equation was utilized for this purpose.

Exact inference on Bayesian networks is widely acknowledged as a computationally challenging task, characterized by its \#P-hard complexity, thus practitioners usually resort to approximate inference techniques when dealing with such networks. These techniques are employed to draw samples from the distribution concerning query variables, given the provided evidence variable values $(e)$. Through the implementation of a quantum adaptation of rejection sampling, a substantial enhancement in efficiency is achieved. For a Bayesian network containing $n$ variables, each with at most $m$ parents per node, a single unbiased sample is obtained with classical resources in time $O(n m P^{-1}(e))$, whereas quantum technologies allow for a square-root speedup to $O(n 2^m P^{-1/2}(e))$ time per sample~\cite{low2014quantum}.

In the paper~\cite{schuld2016prediction} an algorithm for QPU-based prediction was presented, centered around a linear regression model with least-squares optimization. The scheme concentrated on the machine-learning task of predicting the output for a new input based on provided data point examples, and was adapted to handle non-sparse data matrices representable through low-rank approximations, and incorporated substantial enhancements were made to reduce its dependency on the condition number. The prediction's outcome could be obtained through a single-qubit measurement or harnessed for further quantum information processing tasks. To this end the quantum principal component analysis of~\cite{lloyd2014quantum}, as discussed above, was employed. Another quantum algorithm for fitting a linear regression model to a given data set through the least squares approach returning the optimal parameters in classical form was presented in~\cite{wang2017quantum}. The algorithm, once executed, fully determined the fitted model, allowing for cost-effective predictions on new data; it was able to operate on data sets with non-sparse design matrices. Its runtime was characterized by a polynomial dependence on the logarithm size of the data set, the number of adjustable parameters $d$, the condition number of the design matrix $\kappa$, and the desired precision in the output. It was also established that the polynomial dependencies on $d$ and $\kappa$ were essential, indicating that significant improvements to the algorithm were unattainable. Furthermore, a complementary quantum algorithm was introduced to estimate the quality of the least-squares fit without explicitly computing its parameters.

A Determinantal Point Process (DPP) is a type of stochastic process characterized by a probability distribution, which is expressed as a determinant of a certain function. They occur in quantum physics and random matrix theory, and provide effective algorithms for tasks such as sampling, marginalization, conditioning, and other inference operations, and thus are important for ML~\cite{kulesza2012determinantal,derezinski2021determinantal}. Their quantum versions were discussed in~\cite{kerenidis2022quantum}, where novel QML algorithms based on quantum subspace states, offering advancements in quantum linear algebra are introduced. The work considers 3 algorithms. The first algorithm facilitates quantum determinant sampling, achieving a significant speedup compared to classical methods. The second algorithm focuses on quantum singular value estimation for compound matrices, potentially yielding exponential improvements in efficiency. Lastly, the third algorithm reduces the circuit depth for quantum topological data analysis, enhancing computational efficiency.

Recently, orthogonal NNs have emerged as a novel NN architecture that enforces orthogonality on the weight matrices. The characteristic of orthogonality in the trained model weights is utilized to prevent redundancy in the acquired features~\cite{li2019orthogonal,kerenidis2022classical}. The paper~\cite{landman2022quantum} introduced two related novel quantum methods for NNs aimed at enhancing performance in ML applications. The first method, called quantum orthogonal NN, utilizes a quantum pyramidal circuit to implement orthogonal matrix multiplication, with efficient training algorithms for both classical and quantum hardware. The second method, quantum-assisted NNs, employs QPU for inner product estimation during inference and training of classical NNs. Extensive experiments on medical image classification tasks demonstrate similar accuracy levels between quantum and classical NNs, suggesting the potential usefulness of quantum methods in visual tasks as quantum hardware advances.

Several recent papers have explored the application of QC principles to sentiment analysis in natural language processing. The work~\cite{ganguly2022quantum} proposed a method based on the Lambeq toolkit, while~\cite{lai2023quantum} introduced a quantum-inspired fully complex-valued NN, and~\cite{wang2023quantum} presented a quantum-like implicit sentiment analysis approach using sememes knowledge. Liu et al. (2023)~\cite{liu2023survey} surveyed various quantum-cognitively inspired sentiment analysis models. Additionally, Sharma et al. (2022)~\cite{sharma2022comparative} conducted a comparative study between classical and QML models for sentiment analysis. However, before they become effective for finance, the application of quantum sentiment analysis must wait for further advancements.

A model known as the Hybrid classical-quantum Autoencoder was put forth in~\cite{sakhnenko2022hybrid}. It combined the functions of a PQC placed in the bottleneck of a classical autoencoder (AE). Performance was improved in terms of F1 score, recall, and precision once the PQC was included. The benefits of QC in unsupervised AD are illustrated in~\cite{guo2023quantum}. Using QAE and minimal search, the k-distance neighborhood of each data point is identified. Using a quantum multiply-adder, the local reachability density of every data point is computed in parallel. Grover's search for anomalous data points and amplitude estimation are used in parallel to determine each data point's local outlier factor. The method achieves polynomial speedup in the number of data points and exponential speedup in the dimension compared to its traditional equivalent. The work~\cite{Tscharke2023} presents a semisupervised AD method based on the SVR with quantum kernel reconstruction loss (QSVR). This model is evaluated against a classical autoencoder, a quantum autoencoder, and against an SVR with an RBF kernel. QSVR outperforms all these models, earning the greatest mean AUC over all data sets. The models are thoroughly benchmarked on ten real-world AD data sets and one toy data set. QSVR performs better on $9$ out of $11$ data sets compared to the quantum autoencoder. QML for FD in Phishing URL was analyzed in~\cite{Bangroo2023}.

Adiabatic Quantum Algorithms (AQAs) is a type of quantum algorithm that uses adiabatic evolution, a gradual change in the system’s Hamiltonian, to find the ground state of a problem. The basic idea is to start with a simple Hamiltonian that can be easily prepared and solved, and then gradually change it to the problem Hamiltonian, which encodes the solution to the problem. The system will remain in the ground state throughout the evolution if the change is slow enough, and the final state will be the solution to the problem~\cite{albash2018adiabatic}.
Counterdiabatic (CD) driving is a technique used to speed up AQAs. The concept of CD driving was first introduced in physical chemistry in 2003~\cite{demirplak2003adiabatic} and reintroduced in~\cite{berry2009transitionless}. The key idea behind CD driving is to introduce additional terms in the Hamiltonian of the system that counteract the unwanted transitions. By carefully designing these counterdiabatic terms, it is possible to drive the system along a specific path that enhances the evolution towards the desired final state~\cite{Barone_2024}. CD driving has been applied e.g. in QA~\cite{passarelli2020counterdiabatic}.
%
A review of the basic results for quantum error mitigation techniques in hybrid quantum-classical algorithms is given in~\cite{endo2021hybrid}. The issue of implementing QAOA in near-term devices is discussed in~\cite{zhou2020quantum}. 


\subsection{Quantum Machine Learning}
\label{sec:QML}
Quantum machine learning (QML) has emerged as a novel paradigm at the intersection of quantum computing and machine intelligence, aiming to enhance computation speed-up within the current NISQ era using both quantum and classical computational systems and algorithms \cite{wittek2014quantum, biamonte2017quantum, schuld2021machine}. Alas, at the present stage the broader field of quantum machine intelligence \cite{dipierro2022quantum} aiming to develop a learning theory in the quantum domain has proficiently focused on the subfield of QML.  In this area, researchers operate at the frontiers of the available quantum hardware and classical ML, even if the first milestones in QML beared in mind the architecture of a fault-tolerant quantum computer and of a Quantum random-access memory \cite{harrow2009quantum}. Aiming at a comprehensive review of all the relevant contributions in the field is still a Sisyphean endeavor due to the escalating number of publications and patents. However, it signals a growing scientific and industrial interest.  
Therefore, in Section \ref{sec:QML} we briefly review the algorithms that are the most promising for application in quantitative finance, in our opinion. 


Several attempts to translate learning models resembling NNs into the quantum domain, with the primary objective to have a cost function of the form 
\begin{equation}
    \mathcal{L}(\mathbf{x}) = \phi(W \mathbf{x} + b) 
\end{equation}
that could be linked to the unitary transformations that is the core of a quantum circuit and aiming to harness the benefits of quantum information processing.

However, recently the concept of quantum neural networks (QNNs) encompass an entire class of QML algorithms that are based on a combination of parametrized quantum circuits, that 
are subsequently optimized or trained with the aid of a classical processing unit. 
In particular, the link with the NN terms used in classical computer science 
stems from the fact that in the quantum case the hidden layers of the network are composed by the 
sequential action of a given ensemble of parametric gates. 
The main building blocks of a QNN are a preparation routine or feature map \cite{Schuld_PRA2020} in which the data are encoded in a quantum state, the second part of the circuit is the proper variational model containing the gates that are eventually optimised to fulfil the required learning task. The procedure follows the quest of minimizing a loss function minimisation, that is a function of specific measurement outcomes performed on the state. 
In formula, if we denote by $\mathbf{x}$ the data we want to encode in our state then the unitary describing the feature map will be $U_\mathbf{x}$, while the parametric circuit described by the unitary $V(\mathbf{\theta})$ will contain set of trainaible weights $\mathbf{\theta}$ that will be classically optimized. Therefore, the final N-qubit state on which the loss function will be computed reads: 
$V(\mathbf{\theta}) U_{\mathbf{x}}\ket{0}^{\otimes N}.$

Despite the fact that the field is rapidly developing, 
we refer the interest reader to \cite{cerezo2022challenges}, 
where a detailed description of the state of the art with the major challenges facing the quantum machine learning algorithms are thoroughly discussed. 





\section{Literature Review}\label{sec:lit_rev}

The body of literature concerning quantum computing in finance is vast and multifaceted. Notably, several authors, including \cite{albareti, herman2023quantum, orus2019quantum, egger2020quantum, QML_survey4, QML_survey5, pistoia2021quantum}, have presented thorough summaries of the uses of quantum computing in finance. Our literature review, on the other hand, focuses on the use of QML in the finance domain.
While a majority of the sources analyzed entail theoretical explorations employing quantum algorithms for optimization objectives, thus potentially enhancing the speed of various machine learning techniques, only a small subset of these sources present tangible hardware or simulator implementations. Moreover, not all sources contextualize their algorithms within financial applications. Nonetheless, our review encompasses both practical financial implementations and theoretical advancements in quantum machine learning, recognizing the potential applicability of theoretical progress to established finance use cases.
Recently, Wang et al. (2022)~\cite{wang2022design} have proposed a Quantum Finance Software Development Kit (QFSDK) to address the complexities of implementing quantum finance calculations, particularly for educational purposes. This kit, developed in Python, offers a tool for students to gain hands-on experience with quantum finance concepts and applications, thereby enhancing their understanding of this field.

\subsection{Portfolio Optimization and Quantum Machine Learning}
\label{ssec:PortfolioOptimizationAndQuantumMachineLearning}

As discussed in sec.~\ref{sec:intro:portfolio} portfolio optimization is a method that involves selecting the best combination of investments to achieve a specific goal, such as maximizing returns or minimizing risk. 

One of the first QML approaches to portfolio optimization based on quantum annealers was demonstrated in~\cite{rosenberg2015solving}. The authors addressed a multi-period portfolio optimization problem utilizing D-Wave Systems’ quantum annealer, offering a formulation of the problem and discussing various integer encoding schemes. Their numerical examples showcased high success rates, with the formulation accommodating transaction costs and bypassing the need for inverting a covariance matrix. Furthermore, they highlighted the challenges posed by discrete multi-period portfolio optimization and provided insights into potential enhancements for future scalability. For the study~\cite{elsokkary2017financial}, 63 securities listed on the Abu Dhabi Securities Exchange were considered, utilizing the weekly closing value of each over a period of one year. The covariance matrix and expected values were read into the classical CPU with MATLAB along with the budgets and other parameters. The approach tested whether the adoption of the D-Wave QPU allowed for a meaningful increment in computational performance for solving the Markowitz portfolio. D-Wave’s quantum optimizer was used to find the optimal allocation of funds. The D-Wave Solver API (SAPI) embedding routine was used to program the dense Ising model into the unique hardware connectivity graph on the D-Wave processor called Chimera.

The authors of~\cite{venturelli2019reverse} endeavored to address the limitation of the work~\cite{rosenberg2015solving}, which focused on portfolios of insufficient size to assess the scalability of the chosen approach with respect to problem size. They generated parametrized samples of portfolio optimization problems based on real financial data statistics. The samples were linked to quadratic binary optimization forms programmable in the analog D-Wave Quantum Annealer 2000Q. The performance was compared with a genetic algorithm approach with the following results. After investigating various options to optimize quantum computation, they discovered that seeding the quantum annealer with a solution candidate found by a greedy local search and then employing a reverse annealing protocol yielded the best results in terms of expected time-to-solution as a function of the number of variables for the most challenging instance set; this approach was called the optimized reverse annealing protocol. The authors found the method to be more than $100$ times faster on average than the corresponding forward quantum annealing.

In the work~\cite{rebentrost2018quantum} by leveraging quantum access to historical return records, the algorithm determines the optimal risk-return tradeoff curve, facilitating the sampling of the optimal portfolio. If the pertinent data is stored in quantum random access memory, leveraging HHL, quantum walk Hamiltonian simulation methods, and the quantum state exponentiation method~\cite{lloyd2014quantum,kimmel2017hamiltonian}, the matrix pseudo-inverse and quadratic optimization problem can be resolved, potentially achieving a runtime polynomial in $\log(N)$, instead of the classical complexity requiring polynomial time in $N$. Consequently, this approach enables the determination of the risk-return curve and the unveiling of the quantum state associated with the optimal portfolio. To be more specific, this method concerns the unconstrained portfolio optimization problem. The calculations were performed on a QPU emulator. A work appeared soon afterwards~\cite{kerenidis2019quantum} dealt with the constrained portfolio optimization problem with quantum interior point method for second order cone programs~\cite{kerenidis2021quantum} generalized to SDP~\cite{kerenidis2020quantum}. The method was evaluated with a QPU emulator with Gaussian noise with a dataset containing historical data about the stocks of the S\&P-500 companies from the years 2007-2016.

Next, the work~\cite{hodson2019portfolio} applied to the discrete portfolio optimization problem, a different approach than quantum annealing, namely the gate quantum computing model. They evaluate a portfolio rebalancing use case on an idealized simulator of a gate-model QPU, considering characteristics such as trading in discrete lots, non-linear trading costs, and investment constraints. The authors design a novel problem encoding and hard constraint mixers for the Quantum Alternating Operator Ansatz~\cite{hadfield2019quantum} and compare it to the QAOA. Experimental findings indicate that this application is feasible for NISQ hardware, as it can identify portfolios with adjusted returns within 5\% of optimal and optimal risk for a small eight-stock portfolio.

Now let us focus on the recent development of the QA approaches to portfolio optimization. The study~\cite{grant2021benchmarking} focuses on benchmarking on portfolio optimization of the QA controls available in the programmable quantum annealer D-Wave 2000Q, including controls for mapping the logical problem onto hardware and scheduling the annealing process. They explore how the controls influence computational performance and error mechanisms by tuning the quantum dynamics. The study evaluates both forward and reverse annealing methods, identifying control variations that optimize performance e.g. in terms of probability of success. In~\cite{phillipson2021portfolio} the portfolio optimization over D-Wave is compared with conventional commercial solvers, demonstrating that the QA approach shows promising performance, coming close to the performance of existing solvers for problems of similar size. In~\cite{rubio2022portfolio} a quantum-inspired integer simulated annealing method for portfolio optimization in the presence of discretized convex and non-convex cost functions is presented, yet not executed on a quantum hardware. The paper~\cite{lang2022strategic} compares a workflow combining classical preprocessing with modified QUBO models, evaluated on various annealing platforms, including D-Wave, using real-world stock data. The outcomes from QA show promise, although they fell short of the performance achieved with simulated annealing and digital annealing. This discrepancy may be attributed to various factors such as inherent noise, lack of error correction, or scaling issues. In work~\cite{hegade2022portfolio} portfolio optimization using digitized-counterdiabatic quantum computing is explored. This concept is applied to discrete mean-variance portfolio optimization, demonstrating improved success probabilities compared to variational quantum algorithms like QAOA and DC-QAOA. The study highlights the potential of digitized-counter-diabatic quantum algorithms for finance applications in the NISQ era.

The series of papers by Cohen et al.~\cite{cohen2020portfolio,cohen2020portfolio2,cohen2020picking} explores the application of classical and quantum algorithms for portfolio optimization based on the Sharpe ratio, a simplified Chicago Quantum Ratio (CQR), then a new Chicago Quantum Net Score (CQNS), and using U.S. equities. In the first paper~\cite{cohen2020portfolio}, they investigate portfolio optimization of $40$ stocks using the D-Wave Quantum Annealer, exploring various problem formulations based on risk vs return metrics. The second paper~\cite{cohen2020portfolio2} extends this investigation to 60 stocks. Finally, in the third paper~\cite{cohen2020picking}, the authors analyze $3,171$ U.S. common stocks to create efficient portfolios, incorporating classical solvers and QA techniques. These papers collectively demonstrate the potential of both classical and quantum methods for selecting attractive portfolios in the financial domain.

In~\cite{mugel2021hybrid} a dynamic portfolio optimization with a minimal holding period is proposed and was performed on D-Wave 2000Q. The algorithm efficiently samples near-optimal portfolios at each trading step and post-selects to meet the minimal holding constraint. Results indicate that the method produces investment trajectories much closer to the efficient frontier than typical portfolios and can easily adapt to different risk profiles. The work~\cite{palmer2021quantum} demonstrates how to obtain the best investment portfolio with a given target risk and implement individual investment bands (i.e. minimum and maximum possible investments) for each asset to impose diversification and avoid corner solutions. The study utilizes D-Wave Hybrid and its \textit{Advantage} QPU to find optimal portfolios for assets from S\&P100 and S\&P500, showing how practical daily constraints in quantitative finance can be implemented with real data under realistic market conditions. More complex indexes like the Nasdaq Composite can be analyzed with the aid of clustering algorithms. In~\cite{mugel2022dynamic}, the problem of dynamic portfolio optimization over a period of time is addressed using classical solvers, D-Wave Hybrid QA, VQEs on IBM-Q, and a quantum-inspired optimizer based on Tensor Networks. The comparison is taken on real data from daily prices over 8 years of 52 assets. Results indicate that D-Wave Hybrid and Tensor Networks can handle the largest systems effectively.

The paper~\cite{mattesi2023financial} proposes a novel QUBO formulation for portfolio optimization, incorporating both the Sharpe ratio and a diversification term. It modifies the all-or-nothing selection approach of each asset as in~\cite{venturelli2019reverse} by introducing in portfolio construction a linear combination of investments on all assets with arbitrarily large precision. Furthermore, a diversification term is introduced to promote investments across multiple sectors, enhancing portfolio resilience. Results obtained using classical QUBO solvers and the D-Wave Leap hybrid classical-quantum solver demonstrate the effectiveness of the proposed approach in maximizing returns while minimizing sector-specific risks. Optimal outcomes are achieved through D-Wave Leap Hybrid in one of the considered scenarios and classical solver in the other one. The study~\cite{osaba2023quantum} introduces a system, Q4FuturePOP, designed for portfolio optimization with future asset values executed on D-Wave \textit{Advantage System 6.2}, with $5610$ qubits and $40134$ couplers spread over a Pegasus topology for QUBO. Unlike traditional approaches using historical data, Q4FuturePOP utilizes future asset predictions for formulating the optimization problem. Through preliminary evaluations it demonstrates promising performance, surpassing expert solutions from financial advisors like Welzia Management in some instances.

Now, we move to other QML implementation approaches. The study~\cite{alcazar2020classical} compares classical ML, specifically restricted Boltzmann machines (RBMs), with variants of quantum circuit Born machines (QCBMs) implemented on ion-trap QPUs, for a probabilistic version out of the portfolio optimization problem. It utilizes time-series pricing data from asset subsets of the S\&P500 stock market index to assess the performance of both models. The quantum models demonstrated superior performance compared to RBMs when considering the same number of parameters. The effectiveness of certain HHL enhancements is empirically demonstrated through the application to small portfolio optimization problems, which were executed end-to-end on the Quantinuum System Model H1-2 trapped-ion QPU, in work~\cite{yalovetzky2021nisq}. In~\cite{lim2022quantum} a quantum version of an existing classical online portfolio optimization algorithm~\cite{helmbold1998line}, leveraging quantum state preparation, inner product estimation, and multi-sampling techniques, is introduced. The quantum algorithm employs quantum maximum finding, and exhibits a quadratic speedup in time complexity relative to the number of assets in the portfolio $n$, while the transaction cost remains constant in $n$, making it particularly suitable for practical applications with a large number of assets. The authors systematically detail the transition from the classical to the quantum algorithm, starting from an extended version of the classical algorithm, incorporating a sampling procedure to render the transaction cost independent of $n$, and ultimately employing quantum inner product estimation and quantum multi-sampling techniques to devise the quantum online portfolio optimization algorithm.

In~\cite{carrascal2023backtesting} a VQE in solving portfolio optimization problems using simulators and real QPUs with over 100 qubits provided by IBM is analyzed. Comparisons are drawn with three classical algorithms by backtesting. Findings indicate that quantum algorithms exhibit competitiveness with classical counterparts, with the advantage of efficiently handling a large number of assets on future larger QPUs. The paper~\cite{buonaiuto2023best} employs VQE to tackle portfolio optimization efficiently and defines optimal hyperparameters for VQE execution on actual IBM QPUs. By converting the problem into QUBO, with constraints integrated into the objective function, the study identifies key hyperparameters like ansatzes and optimization methods. Through experiments on simulators and real QPUs, the research demonstrates a strong correlation between solution quality and quantum hardware size. It concludes that with proper hyperparameters and sufficiently sized quantum hardware, VQE can produce solutions close to exact ones, even without error-mitigation techniques, suggesting a promising avenue for future optimization endeavors.

\subsection{Market Prediction and Trading using Quantum Machine Learning}
\label{ssec:MarketPredictionAndTradingStrategiesUsingQuantumMachineLearning}

A significant challenge in the financial industry revolves around trading and hedging portfolios of derivatives. Recently, an innovative approach has been introduced to address this issue without relying on frictionless and complete market assumptions \cite{buehler2019deep}. In this approach, trading decisions within hedging strategies are modeled as NNs in a reinforcement learning framework. Expanding on this work, \cite{cherrat2023quantum} adapts the problem studied by \cite{buehler2019deep} to a quantum-native setup. In this quantum framework, market states are encoded into quantum states, and policies and value functions are represented using Quantum Neural Networks (QNNs). The implementation is carried out on the 20-qubit trapped-ion quantum processor, Quantinuum H1, with training conducted through gradient descent. The effectiveness of this quantum approach is compared against the Black-Scholes delta hedge model, and the results show that the QNN policies outperform the traditional model significantly.

The research conducted in~\cite{von2021quantum} delves into metrics utilized for encoding combinatorial search as a binary quadratic model tailored for feature selection. The primary objective is to enhance the model's overall performance, particularly in terms of generalization, model fit, and accuracy when applied to the regression task of price prediction. Through the utilization of quantum-assisted routines with QUBO using the D-Wave Advantage 1.1 sampler, the authors observed notable enhancements in the quality of predictive model outputs, coupled with a reduction in input dimensionality for the learning algorithm across synthetic and real-world datasets.

In~\cite{liu2022quantum} the so-called quantum Elman NN (QENN) was investigated. The Elman NN (ENN) itself, is a type of recurrent NN developed by Jeffrey Elman in 1990~\cite{elman1990finding} to process sequential data and has been widely used in various fields including time series prediction. The ENN consists of an input layer, a hidden layer, a context layer, and an output layer. The input layer receives the classical input data, which is then processed through the hidden layer, which in the quantum case can be a quantum register. The context layer, which can also be a quantum register, stores the previous state of the hidden layer and provides feedback to the network, allowing it to remember past information and learn temporal dependencies in the data. Finally, the output layer produces the network’s output, possibly in quantum form, based on the processed information.
The learning rates in~\cite{liu2022quantum} are tuned using a quantum version of genetic algorithms. The method was applied for the prediction of closing prices on the Nasdaq, BSE Sensex, HSI, SSE, Russell 2000, and TAIEX stock markets. It was shown that QENN can attain the quality of prediction of ENN with the much smaller size of the hidden and context layers.

The study~\cite{paquet2022quantumleap} unveils a novel hybrid deep QNN designed for financial forecasting tasks. Central to the approach is an encoder module that converts partitioned financial time series into a sequence of density matrices, followed by the utilization of a deep quantum network to forecast the density matrix at a future time step. The research demonstrates that the maximum price attained by security at a later time can be extracted from the output density matrix. Through extensive experimentation involving $24$ securities, the system showcases remarkable accuracy and efficiency across both regression and extrapolation scenarios.

The paper~\cite{daskin2022walk} examined the quantum analogs of classical data preprocessing and forecasting utilizing autoregressive integrated moving average (ARIMA) time series models, employing straightforward quantum operators with minimal quantum gate requirements. The study~\cite{emmanoulopoulos2022quantum} also explored time series forecasting using quantum technologies. They investigated the effectiveness of PQCs employed as QNN for predicting time series signals through simulated quantum forward propagation. Their findings suggest that QNNs can proficiently model time series data while offering the notable advantage of faster training compared to classical machine learning models when executed on QPUs. Another study~\cite{rivera2022time} focusing on time series forecasting introduces two classical-quantum hybrid architectures employing QNNs and Hybrid-Quantum Neural Networks. These architectures integrate quantum variational circuits with specialized encoding schemes, with optimization executed by a classical computer. The experiments were performed on an QPU emulator. Performance validation across four representative forecasting problems (e.g. USD-to-EUR currency exchange rate) demonstrates competitive performance, despite a comparable number of trainable parameters relative to classical solutions.

The study~\cite{thakkar2023improved} demonstrates two applications within Ita\'{u} Unibanco, Latin America's largest bank. Quantum algorithms for DPP were applied to enhance Random Forest models for churn prediction. These QML algorithms improved precision by nearly 6\% compared to baseline models. So-called quantum compound NN architectures which utilize quantum orthogonal NNs, were designed for credit risk assessment. The experiments were conducted on IBM 16-qubits platform \textit{guadelupe}. These NN architectures demonstrated superior accuracy and generalization compared to classical fully-connected NNs while requiring fewer parameters.

In~\cite{Srivastava2023} QUBO on D-Wave was employed for feature selection, and Principal Component Analysis (PCA) was used for dimensionality reduction. The task of stock price prediction was transformed into a classification problem, and the QSVM was trained to predict price movements. The performance of QSVM was compared with classical models, and their accuracy was analyzed using datasets formulated with QA and PCA. The study focused on predicting stock prices and binary classification for four companies: Apple, Visa, Johnson and Johnson, and Honeywell, using real-time stock data. Various Quantum Computing techniques were compared with their classical counterparts in terms of prediction model accuracy and F-score. QA showcased superior efficacy in extracting the most pertinent features from financial data compared to PCA. However, QSVM did not demonstrate a notable advantage over classical SVM with the provided datasets.

\subsection{Pricing and Quantum Machine Learning}
\label{ssec:DerivativePricingQML}

The Heath-Jarrow-Morton (HJM) model, widely employed in finance for valuing interest rate derivatives~\cite{heath1992bond}, encounters a notable challenge due to its extensive degrees of freedom when describing the evolution of the yield curve. One potential strategy to tackle this challenge involves the application of principal component analysis for factor selection. The use of quantum Principal Component Analysis (qPCA) can effectively reduce the number of noisy factors as shown in \cite{martin2021toward}, facilitating the determination of fair prices for interest rate derivatives. The estimation of principal components for $2 \times 2$ and $3 \times 3$ cross-correlation matrices, based on historical data for two and three time-maturing forward rates, is executed using the 5-qubit IBMQX2 quantum processor. The results indicate that the algorithm can provide reasonable approximations for the $2 \times 2$ case, although the quantum processor faces limitations related to gate fidelities, connectivity, and the number of qubits.
Simultaneously, experimental outcomes with simulators suggest that improved results could be achievable with the availability of a lower-level programming interface. Such an interface would enable the customization of quantum algorithm optimization to align with chip constraints, offering a promising avenue for refinement.

In classical machine learning, Generative Adversarial Networks (GANs) excel at generative modeling, an unsupervised learning approach involving a generator and a discriminator engaged in a competitive training process. The introduction of quantum systems, replacing the generator, discriminator, or both, extends this framework into the domain of quantum computing. An exemplary application is demonstrated in the work~\cite{zoufal2019quantum}, where quantum-classical hybrid GANs are employed to learn and transfer approximations of probability distributions from classical data to gate-based QPUs. This is an efficient, approximate data loading scheme that requires significantly fewer gates than existing methods. Specifically, a log-normal distribution is learned that models the spot price of an underlying asset for a European call option. Finally, QAE is used to estimate the expected payoff of the option, given the efficient, approximate data loading by the quantum GANs (qGANs). The training and loading are run on an actual QPU, the IBM Q Boeblingen chip with 20 qubits, with the gradient-based optimization of the qGAN parameters taking place on a classical computer.

The earliest quantum attempts at derivative pricing were presented in two papers by Chen~\cite{chen2001quantum1} and Chen~\cite{chen2001quantum} in 2001. In~\cite{chen2001quantum1} a quantum adaptation of select areas of arbitrage theory, asset pricing, and optional decomposition within financial markets, were presented, operating on finite-dimensional quantum probability spaces. The work resolved certain paradoxes in classical models and re-deriving option pricing formulas. In~\cite{chen2001quantum} a quantum model for binomial markets was proposed, preventing arbitrage opportunities and revisiting option pricing formulas from a quantum perspective. In~\cite{rosenberg2016finding} two formulations for finding optimal arbitrage opportunities as a QUBO solved using D-Wave were presented.

An early quantum algorithm for pricing of financial derivatives was presented in~\cite{rebentrost2018quantumDerivatives}. The relevant probability distributions were prepared in quantum superposition, and the payoff functions were implemented via quantum circuits, with the price of financial derivatives extracted via quantum measurements. QAE on an emulator was applied to achieve a quadratic quantum speedup in the number of steps required to obtain an estimate for the price with high confidence. This study combined QAE and the quantum algorithm for Monte Carlo, with the pricing of financial derivatives, and provided a foundation for further research at the intersection of quantum computing and finance.

The paper~\cite{fontanela2021quantum} introduced a hybrid quantum-classical algorithm, inspired by quantum chemistry, for pricing European and Asian options in the Black-Scholes model. By leveraging the equivalence between the pricing PDE and the Schroedinger equation in imaginary time, the algorithm transforms the Black-Scholes PDE into the Heat Equation and represents the option price as a wave function. This wave function is then solved using a hybrid quantum and classical algorithm, incorporating McLachlan's invariance principle~\cite{mclachlan1964variational} to build a quantum circuit of imaginary time evolution. Despite requiring only a few qubits on an emulator, the shallow quantum circuit accurately represented European and Asian call option prices, indicating a promising potential for applying quantum computing techniques in quantitative finance.

The paper~\cite{ramos2021quantum} introduced a quantum algorithm for European option pricing in finance, employing a unary representation of the asset value. The algorithm first generates the amplitude distribution corresponding to the asset value at maturity using a low-depth circuit; then it computes the expected return with simple controlled gates; and finally employs the QAE. A comparison of unary and binary option pricing algorithms executed on Qiskit emulator and using error maps indicated that unary representation could offer a significant advantage in practice for near-term devices.

In~\cite{stamatopoulos2020option} pricing of various types of options, including vanilla options, multi-asset options, and path-dependent options such as barrier options, was examined using the gate-based IBM Q Tokyo quantum device showing a quadratic speed-up compared to traditional Monte Carlo simulations. Complex features in exotic options, such as path dependency with barriers and averages, were addressed. The results rely on QAE, and its variant without phase estimation~\cite{suzuki2020amplitude} was demonstrated to reduce the number of gates required for measuring option prices. An effective error mitigation scheme was employed to reduce errors arising from noisy two-qubit gates.

In~\cite{doriguello2021quantum}, a quantum least squares Monte Carlo (LSM) approach is introduced that leverages quantum access to a stochastic process, utilizes quantum circuits for computing optimal stopping times, and employs quantum techniques for Monte Carlo simulations. A nearly quadratic speedup in runtime compared to traditional LSM methods is demonstrated, and examples of its application to American option pricing are given. In~\cite{miyamoto2022bermudan} LSM was applied to Bermudan option pricing. This method approximates the continuation value, a crucial component of Bermudan option pricing, through Chebyshev interpolation. It utilizes values at interpolation nodes estimated by QAE and demonstrates a quadratic speed-up compared to classical LSM. In~\cite{radha2021quantum} the preparation of an initial state representing the option price, followed by its evolution using existing time simulation algorithms in Wick’s imaginary time-space were employed for pricing options. Due to its utilization of a hybrid variational algorithm, the proposed method was deemed relevant for NISQ QPUs. The method was numerically verified for European options and has potential extensions to path-dependent options like Asian options.

In~\cite{miyamoto2021pricing}, the authors focus on derivative pricing through the solution of the Black-Scholes partial differential equation using the finite difference method (FDM), a suitable approach for certain types of derivatives but challenged by the so-called curse of dimensionality. They introduce a quantum algorithm for FDM-based pricing of multi-asset derivatives, demonstrating an exponential speedup in dimensionality compared to classical algorithms. Leveraging quantum differential equations solving algorithms~\cite{berry2017quantum}, the proposed approach addresses the issue of extracting derivative prices from the output state of the quantum algorithm, outlining the calculation process and estimating its complexity. The work optionally used also QAE for the calculation of the so-called nodal continuation values important for Bermudan option pricing. In the subsequent study~\cite{kubo2023pricing}, the same authors make their algorithm feasible to run on a small QPU but by avoiding certain efficiency bottlenecks of embedding derivative price in the amplitude, they use variational quantum simulation to solve the Black-Scholes equation and compute the derivative price from the inner product between the solution and a probability distribution. They employed a QPU emulator. In~\cite{li2023quantum}, a quantum Monte Carlo algorithm was proposed to solve high-dimensional Black-Scholes PDEs with correlation and general continuous and piece-wise affine payoff functions. The approach involves uploading the multivariate log-normal distribution and the rotated form of the payoff function; subsequently, QAE is applied. Error and complexity analyses show that the computational complexity grows only polynomially in the space dimension of the Black-Scholes PDE and the reciprocal of the accuracy level, indicating that the algorithm is not afflicted by the curse of dimensionality. The results were verified on a QPU emulator \textit{qfinance} using Qiskit.

In~\cite{chakrabarti2021threshold} a strategy to optimize the PQC for pricing a specific type of derivatives, known as Target Accrual Redemption Forward, combining elements of an option and a forward contract, was introduced. This strategy was based on an energy-based method proposed in~\cite{peruzzo2014variational}. It combined pre-trained variational circuits with fault-tolerant quantum computing to reduce resource requirements. The target cost function was determined as the energy of the associated quantum harmonic oscillator problem, Gaussian in its ground state~\cite{ollitrault2020nonadiabatic}. The circuits required to encode these states for different choices of the register size $n$ were pre-trained and applicable to any derivative pricing problem, thereby eliminating the need to include training costs in overall resource estimations. The numerical study demonstrated that the state prepared variably approached the target exponentially fast in terms of the number of gate operations. In the study~\cite{kaneko2022quantum} QAE is the primary source of quantum speedup in a model in which the volatility of the underlying asset price depends on the price and time. The paper explores two variants for the state preparation step of QAE: amplitude encoding, where the probability distribution of the derivative's payoff is encoded into probabilistic amplitudes, and pseudo-random number (PRN) type, where sequences of PRNs simulate asset price evolution akin to classical Monte Carlo simulation.

Other noticeable quantum approaches to option pricing, which we yet not classify as QML include~\cite{tang2021quantum,an2021quantum,gonzalez2023efficient,yecsiltacs2023black}. A so-called quantum first fundamental theorem of asset pricing stating the equivalence between no-arbitrage and the existence of a risk-free density operator under which all assets are martingale was given in~\cite{bao2023fundamental}.

\subsection{Risk Management and Quantum Machine Learning}
\label{ssec:RiskAssessmentUnderwritingQML}

In a pioneering work~\cite{woerner2019quantum}, a quantum algorithm has been presented to analyze risk more efficiently compared to traditional Monte Carlo simulations on classical computers. QAE was utilized to price securities and assess VaR and Conditional VaR on a gate-based QPU. The implementation of this algorithm and the trade-off between convergence rate and circuit depth were demonstrated, indicating a near quadratic speed-up compared to Monte Carlo methods as circuit depths increase gradually. Two toy models were employed to demonstrate the algorithm's efficacy. The first model utilized real hardware, the IBM Q Experience, to price Treasury bills, representing short-term debt obligations, under the risk of potential interest rate increases. The second model simulated the algorithm to showcase how a QPU can evaluate financial risk for a two-asset portfolio comprising government debt with varying maturity dates. Both models confirmed the improved convergence rate over Monte Carlo methods.

In paper~\cite{dasgupta2019quantum}, a QA algorithm in QUBO form for a dynamic asset allocation problem with an expected shortfall constraint is presented. The algorithm, which is dynamic and allows the risk target to emerge from the market volatility, is formulated in a manner suitable for implementation by a quantum annealer like D-Wave.

The work~\cite{egger2020credit} introduces a quantum algorithm based on QAE for VaR estimation and evaluates it on a Qiskit QPU emulator. The algorithm was designed for the efficient estimation of credit risk, surpassing the capabilities of classical Monte Carlo simulations. Specifically, the algorithm focuses on estimating the ECR. The study implements this algorithm for a realistic loss distribution, providing a comprehensive analysis of its scalability to practical problem sizes. It offers insights into the total number of required qubits, the anticipated circuit depth, and the expected runtime under reasonable assumptions regarding future fault-tolerant quantum hardware. The conclusions highlight a quadratic speedup achieved by the quantum algorithm in estimating economic capital requirements, supported by a simulation that considers realistic problem sizes. The scalability and expected runtime are thoroughly examined, with the suggestion that the results extend to more intricate uncertainty models or alternative objectives, such as conditional value at risk, with minimal additional computational overhead. The development of a quantum circuit for the Gaussian conditional independence model is detailed, with the acknowledgment that diverse credit risk models would necessitate their dedicated quantum circuits.

In~\cite{herr2021anomaly}, variational quantum-classical Wasserstein GANs were presented to solve the problems of sampling efficiency limits and GAN training instabilities. The model maintained the structure of the classical discriminative model but substituted a QNN for the Wasserstein GAN generator to ensure that there was no need to prepare high-dimensional classical data in a quantum circuit. In terms of F1 score, the effectiveness of a dataset of credit card fraud was comparable to the traditional method. The work examined with TensorFlow Quantum emulator how to sample noise, layer depth and width, sampling noise, and the initialization strategy for QNN design parameters affected convergence and performance.

Feature selection stands as a challenging and crucial task within machine learning, involving the identification of a subset of pertinent features in a dataset from the original set. This process differs from feature extraction, which generates new features from the original set, capturing essential information in a lower-dimensional space. Utilizing QNNs, a quantum algorithm for feature selection is introduced in~\cite{Zoufal_2023}. Specifically, QNNs are trained to generate feature subsets that optimize the performance of a predictive model. While any arbitrary classifier and scoring function could be used, the study opts for logistic regression as the classifier and log-loss as the performance score.
The efficacy of the QNN feature selection method is assessed using a publicly available real-world credit risk dataset containing $1000$ data points that evaluate a customer's creditworthiness based on 20 attributes. The optimizer for training the QNN parameters is a version of the gradient-free method called simultaneous perturbation stochastic approximation. The feature selection algorithm is implemented on superconducting quantum hardware \textit{ibmq montreal} with $27$ qubits, demonstrating results that compete with state-of-the-art classical methods and, in certain experiments, surpass them.

In~\cite{vesely2022application}, it was found that QPUs were able to solve certain tasks related to Foreign Exchange reserves management, including risk measurement using the quantum Monte Carlo method and Markowitz-like portfolio optimization employing the HHL algorithm and QAOA. However, due to current hardware limitations, the application of QAOA for a task with only five binary variables and only a few of the demonstrations were successful. The application of the quantum Monte Carlo method to risk measurement generated only partially correct results, and the use of the HHL algorithm in portfolio optimization failed. Running the algorithms discussed above on a QPU simulator and on IBM Lima superconducting QPU confirmed the correctness of the implementations. The work also provided a very concise yet comprehensive overview of general quantum computation with a tutorial.


After discussing the risk assessment methods we move to the topic of FD with QML. Early promising developments for FD with QML were discussed in~\cite{di2021quantum}, with conclusion that QPU available at the time of the work were not stable and error tolerant for practical problems.
In~\cite{tapia2022fraud}, different single-qubit architectures proposed in~\cite{perez2020data,perez2021one} were analyzed and a novel implementation was presented using the Qiskit. The former trained a single-qubit based on the concept of data re-uploading, allowing the encoding of mathematical functions in the degrees of freedom of a series of gates applied to a single-qubit state. The authors trained a QNN on real data with Qiskit QPU emulator, and benchmarked it against a NN algorithm, particularly using the Kaggle credit card FD dataset with $284807$ transactions, out of which $492$ are fraudulent~\cite{kaggle_credit_card}. Different accuracies associated with various layer formulations, the use of an initial data loading layer, and different numbers of layers for a specific problem were demonstrated.
In~\cite{kyriienko2022unsupervised}, more complex QNN models were used on that dataset. The results established classical benchmarks based on supervised and unsupervised ML methods, with average precision chosen as a robust metric for detecting anomalous data. Quantum kernels of different types were employed for performing AD, and it was observed that the method could challenge equivalent classical protocols as the number of features, equal to the number of qubits for data embedding, increased. Simulations with registers up to $20$ qubits showed that quantum kernels with re-uploading demonstrated better average precision, with the advantage increasing with system size. At $20$ qubits, the quantum-classical separation of average precision was equal to $15\%$.
In~\cite{tekkali2023smart} QML was employed for fraud identification in digital transactional payments using the datase~\cite{kaggle_credit_card}. The implementation of the QNN was emulated using Python, and it was observed that the classical Neural Network required about $4.7$ times more time and achieved lower accuracy ($95.37\%$) compared to QNN.
In~\cite{Wang2023}, a QNN was introduced to learn directly from raw images to train a normality model. It was demonstrated that a quantum-classical hybrid solution , executed on the QPU emulator Rigetti’s Forest SDK, can outperform its classical counterpart, even when they have the same number of learnable parameters. Other works concerning AD not directly related to finance include~\cite{ghosh2023exploring,oh2023quantum}.

A hybrid system integrating quantum and classical machine learning algorithms for the detection of phishing attacks within financial transaction networks based on the Ethereum blockchain is proposed by~\cite{ray2022classical}. Data is accessed through the Etherscan block explorer, a tool for the open-source public blockchain platform Ethereum. Phishing account labels are derived from public reports on phishing activities, resulting in a dataset of 3 million nodes. Among these, 1165 nodes ($0.039\%$) are identified as phishing, creating a high class-imbalance scenario in the classification task. QNNs and QSVMs are employed for this purpose, with extensive testing of QNNs involving various parametrization schemes and QSVMs implemented on both annealers and gate-based devices.
Optimal configurations for the models are determined through simulators, and the study conducts exhaustive experimentation using these optimized models on IBM's 5- and 27-qubit chips, as well as a D-Wave annealer with 5617 qubits. Surprisingly, the results do not indicate a performance improvement with an increased number of qubits. In the optimization of QNNs, the classical optimizer chosen is the gradient-free algorithm known as constrained optimization by linear approximation. The study reveals that stacking and bagging, techniques that capitalize on the complementary strengths of quantum and classical models, lead to improved results.
The findings highlight that gate-based QSVMs consistently yield lower false positives, resulting in higher precision compared to other classical and quantum models. This characteristic is particularly valuable in the context of AD problems.

In~\cite{grossi2022mixed}, an application of a QSVM algorithm, utilizing the IBM Safer Payments and IBM Quantum Computers via the Qiskit software stack and real card payment data, for a classification problem in the financial payment industry was presented. A novel method for searching for the best features was explored using the QSVM’s feature map characteristics. The results were compared with classical solutions using fraud-specific key performance indicators: Accuracy, Recall, and False Positive Rate, extracted from analyses based on human expertise (rule decisions), and classical machine learning algorithms. The QSVM provided a complementary exploration of the feature space, leading to improved accuracy of the mixed quantum-classical method for FD, despite the use of a drastically reduced data set to fit the current state of Quantum Hardware.
The paper~\cite{wang2022integrating} proposed a detection system, which was implemented with SVM supplemented with D-Wave QA. Twelve machine learning methods were further examined to assess their detection performance, and QSVM was contrasted with them on two datasets: a highly imbalanced bank loan dataset~\cite{Mishra50012019-dataset} (time series) and a moderately imbalanced Israel credit card transactions~\cite{Polantizer-dataset} (non-time series). With the former dataset, the QSVM was found to perform better than the others in terms of speed and accuracy; however, with the latter dataset, its detection accuracy was comparable to that of the others. It was demonstrated for both datasets that feature selection greatly increased the detection speed while only slightly increasing the accuracy.
A comparison of four QML models on Qiskit QPU emulator was done in~\cite{Innan2024compare}, and QSVM performed the best, with F1 scores of $0.98$ for both fraud and non-fraud classes. Promising outcomes were also shown by other quantum models.

In~\cite{Guan2022}, a framework was established for QML fairness verification. Adopting the fairness notion which asserts that any two similar individuals must be treated similarly to ensure unbiased treatment, the study explored how quantum noise could potentially enhance fairness. An algorithm was formulated based on Tensor Networks and implemented using Google’s TensorFlow Quantum to ascertain whether a (noisy) QML model adhered to fairness principles. Experimental results, including income prediction and credit scoring based on real-world data, validated the utility and effectiveness of the algorithm, particularly for a class of random (noisy) quantum decision models characterized by 27 qubits (resulting in $227$-dimensional state space). Subsequently, in~\cite{Guan2023}, the authors extended their work by defining a formal framework for detecting violations of differential privacy.

The work~\cite{miyabe2023quantum} involved a hybrid, quantum multiple kernel learning (QMKL) approach aimed at enhancing classification quality compared to a single kernel method. Robustness testing of QMKL was conducted across various financially relevant datasets utilizing both fidelity and projected quantum kernel techniques. Application of the QMKL encompassed multiple financially related datasets, including HSBC Digital Payment data. Both fidelity quantum kernel~\cite{havlivcek2019supervised} and the more recent projected quantum kernel~\cite{huang2021power} techniques underwent testing in simulation and practical demonstration on quantum hardware \textit{ibm\_auckland}. Hardware implementation was optimized using an error mitigation pipeline comprising randomized compiling to mitigate coherent errors and pulse-efficient transpilation to reduce temporal overhead for cross-resonance gates during two-qubit unitary rotations. This pulse transpilation strategy facilitated scaling of the feature space up to $20$ qubits on hardware.
In~\cite{kolle2023towards}, the results from~\cite{kyriienko2022unsupervised} were reassessed and utilized as benchmarks, on a Qiskit emulator, to assess the efficacy of two linear time complexity methods based on data size: randomized measurements for quantum kernel measurement~\cite{haug2023quantum} and an ensemble method termed variable subsampling~\cite{aggarwal2017introduction}. The dataset from~\cite{kaggle_credit_card} was employed for training purposes. It was revealed that while attainable improvements in average precision and F1 score over the classical kernel were observable, they were not deemed very significant. Models utilizing variable subsampling with the inversion test demonstrated stability, whereas those employing the randomized measurement method exhibited high variance. Variable subsampling notably manifested considerable enhancements in training and testing times, suggesting potential performance elevation opportunities through alternate hyperparameters.

In~\cite{leclerc2023financial}, a QML method for fallen-angels forecasting with a quantum-enhanced classifier based on the QBoost algorithm~\cite{neven2008training,neven2009training} was proposed. This solution was implemented on a neutral atom QPU with up to $60$ qubits using a real-life dataset. The proposed classifier, trained on the QPU, achieved competitive performance with $27.9\%$ precision compared to the state-of-the-art Random Forest benchmark, which achieved $28\%$ precision for the same recall of approximately $83\%$. However, the proposed approach outperformed its classical counterpart in terms of interpretability, employing only $50$ learners compared to $1200$ for the Random Forest, while maintaining comparable runtimes.
In~\cite{innan2024financial}, an approach for detecting financial fraud using Quantum GNNs was proposed. Classical and quantum GNNs benchmark using a real-world financial fraud detection dataset showed that the latter outperformed the former and achieved an AUC of $0.85$.

\section{Conclusions}

We have provided a comprehensive overview of the applications of QML in finance. Through an exploration of various use cases including portfolio optimization, market prediction, pricing, and risk management, the potential of QML to improve financial analysis and decision-making has been highlighted. By examining the synergy between quantum computing and machine learning, insights into the future of quantitative finance have been elucidated. Despite the promising advancements and competitive performance demonstrated by QML algorithms, challenges such as scalability, hardware limitations, and algorithmic complexity remain. Addressing these challenges will be crucial for realizing the full potential of QML in finance. Overall, this review underscores the importance of continued research and development in QML for advancing quantitative finance and unlocking new opportunities in the financial industry.

Quantum technologies offer promising applications in portfolio optimization, leveraging quantum computing's potential to efficiently solve complex optimization problems. Techniques such as QA and VQEs have been explored to address portfolio optimization challenges. QA algorithms have been employed to find optimal portfolios by minimizing risk while maximizing returns, while VQE algorithms provide a quantum approach to compute the eigenvalues of portfolio matrices. These quantum approaches aim to enhance the efficiency and accuracy of portfolio optimization, potentially outperforming classical optimization methods as quantum computing hardware continues to advance. We also note that in certain cases quantitative finance inspires quantum methods, e.g. in~\cite{barkoutsos2020improving,chen2023quantum} Conditional Value-at-Risk concept was utilized to enhance the efficiency of general quantum optimization techniques.

\section*{Acknowledgements}

This work was initiated when PM was partially, ASH and AM were fully supported by the Foundation for Polish Science (IRAP project, ICTQT, contract No. 2018/MAB/5, co-financed by EU within Smart Growth Operational Programme). PM also acknowledges support from the Knut and Alice Wallenberg Foundation through the Wallenberg Centre for Quantum Technology (WACQT), and from NCBiR QUANTERA/2/2020 (www.quantera.eu) an ERA-Net cofund in Quantum Technologies under the project eDICT. The work of EY and TA was carried out as part of the IFZ FinTech program with financial support from various industry partners and the Lucerne University of Applied Sciences and Arts.

\bibliographystyle{ACM-Reference-Format}
\bibliography{references}


\end{document}